\pdfoutput=1
\documentclass{tlp}
\usepackage{graphicx}
\usepackage[T1]{fontenc}
\usepackage{xspace}
\usepackage{thmtools}
\usepackage{fancybox}
\usepackage{times}
\usepackage{url}
\usepackage{amsmath}
\usepackage{amssymb}
\usepackage{graphics}
\usepackage{amsfonts}
\usepackage{algorithm}
\usepackage{algorithmicx}
\usepackage{algpseudocode}
\usepackage{subcaption}
\usepackage{multirow}
\usepackage{listings}
\usepackage{color}

\newtheorem{definition}{Definition} 
\newtheorem{example}{Example} 

\def\inv{\vspace{-0.2cm}}
\newcommand{\stitle}[1]{\vspace{0.3ex} \noindent{\bf #1}}

\newcommand{\ctab}{Table~}

\newcommand{\cfig}{Figure~}

\newcommand{\cexp}{Example~}

\newcommand{\at}[1]{\protect\ensuremath{\mathsf{#1}}\xspace}

\def\mt{\tt}
\newcommand{\arrow}{~\mbox{\tt <-}~}
\newcommand{\bldl}{\[\begin{array}{ll}}
\newcommand{\cldl}{\vspace*{-0.3cm}\[\begin{array}{ll}}
\newcommand{\eldl}{\end{array}\]}
\newcommand{\prule}[2]{& \mt #1 \arrow \mt #2 \\}

\newcommand{\pbody}[2]{& \mt #1 \mt #2 \\}

\newcommand{\maxv}[1]{{\sf max} \langle {\tt #1} \rangle}
\newcommand{\minv}[1]{{\sf min} \langle {\tt #1} \rangle}
\newcommand{\cntv}[1]{{\sf count} \langle {\tt #1} \rangle}

\newcommand{\sumv}[1]{{\sf sum} \langle {\tt #1} \rangle}

\def\mminv#1{mmin\langle #1 \rangle}

\def\maxv#1{max\langle #1 \rangle}
\def\minv#1{min\langle #1 \rangle}
\def\sumv#1{sum\langle #1 \rangle}

\def\countv#1{count\langle #1 \rangle}

\def\prem{$\cal P$$\!reM$\xspace}

\newcommand{\bigdatalog}{\at{BigDatalog}}
\newcommand{\bigdatalogmc}{\at{BigDatalog\!\!-\!\!MC}}
\newcommand{\bigdatalogspark}{\at{BigDatalog\!\!-\!\!Spark}}

\newcommand{\ornode}[1]{\mbox{\sc Or-{#1}}}
\newcommand{\andnode}[1]{\mbox{\sc And-{#1}}}
\newcommand{\gettuple}[1]{{#1}.{\tt getTuple}}

\newcommand{\bdealrtl}{\small\smallskip\[\begin{array}{@{}l@{\hspace{0.05cm}}@{}l@{\hspace{0.05cm}}}}
\newcommand{\edealrtl}{\end{array}\]\normalsize}
\newcommand{\prulertlt}[3]{\mt \it{#1.} & \mt #2 \leftarrow #3 \\}

\def\tc{\at{TC}}
\def\sg{\at{SG}}

\def\sssp{\at{SSSP}}
\def\cc{\at{CC}}
\def\attend{\at{ATTEND}}
\def\bfs{\at{REACH}}
\def\pymk{\at{PYMK}}
\def\mlm{\at{MLM}}
\def\psn{Parallel Semi-na{\"i}ve}
\def\rone{$\tt r1$}
\def\rtwo{$\tt r2$}
\def\rthree{$\tt r3$}

\def\mmax{{\tt mmax}}
\def\mmin{{\tt mmin}}
\def\mcnt{{\tt mcount}}
\def\msum{{\tt msum}}

\def\treeeleven{{\tt Tree11}}
\def\treeseventeen{{\tt Tree17}}
\def\gridone{{\tt Grid150}}
\def\gridtwo{{\tt Grid250}}
\def\gfivek{{\tt G5K}}
\def\gtenk{{\tt G10K}}
\def\gtenkzeroone{{\tt G10K-0.01}}
\def\gtenkone{{\tt G10K-0.1}}
\def\gtwentyk{{\tt G20K}}
\def\gfourtyk{{\tt G40K}}
\def\geightyk{{\tt G80K}}

\def\figurename{Figure}
\def\examplename{Program}

\algblock{ParFor}{EndParFor}
\algblock{DoWhile}{EndDoWhile}
\algnewcommand\algorithmicparfor{\textbf{parallel for}}
\algnewcommand\algorithmicpardo{\textbf{do}}
\algnewcommand\algorithmicendparfor{\textbf{end\ parallel for}}
\algrenewtext{ParFor}[1]{\algorithmicparfor\ #1\ \algorithmicpardo}
\algrenewtext{EndParFor}{\algorithmicendparfor}

\algnewcommand\algorithmicdowhile{\textbf{do}}
\algnewcommand\algorithmicenddowhile{\textbf{while}}
\algrenewtext{DoWhile}{\algorithmicdowhile}%
\algrenewtext{EndDoWhile}[1]{\algorithmicenddowhile\ #1}

\begin{document}
\bibliographystyle{acmtrans}

\long\def\comment#1{}

\title{Scaling-Up Reasoning and Advanced Analytics on BigData}
\author[Condie et Al.]{
Tyson Condie,$\;$Ariyam Das,$\;$Matteo Interlandi,$\;$Alexander Shkapsky,$\;$Mohan Yang,$\;$Carlo Zaniolo\\
University of California, Los Angeles\inv\\\\
\email{\small  \{tcondie,ariyam,minterlandi,shkapsky,yang,zaniolo\}@cs.ucla.edu}
}

\pagerange{\pageref{firstpage}--\pageref{lastpage}}
\volume{\textbf{10} (3):}
\jdate{March 2002}
\setcounter{page}{1}
\pubyear{2002}

\maketitle  

\label{firstpage}

\begin{abstract}

BigDatalog is an extension of Datalog that achieves performance and scalability on both Apache Spark and multicore systems to the point that its graph analytics outperform those written in GraphX. Looking back, we see how this realizes the ambitious goal pursued by deductive database researchers beginning forty years ago: this is the goal of combining the rigor and power of logic in expressing queries and reasoning with the performance and scalability by which relational databases managed Big Data. This goal led to Datalog which is based on Horn Clauses like Prolog but employs implementation techniques, such as Semi-na{\"i}ve Fixpoint and Magic Sets, that extend the bottom-up computation model of relational systems, and thus obtain the performance and scalability that relational systems had achieved, as far back as the 80s, using data-parallelization on shared-nothing architectures. But this goal proved difficult to achieve because of major issues at (i) the language level and (ii) at the system level. The paper describes how (i) was addressed by simple rules under which the fixpoint semantics extends to programs using count, sum and extrema in recursion, and (ii) was tamed by parallel compilation techniques that achieve scalability on multicore systems and Apache Spark.
This
paper is under consideration for acceptance in Theory and Practice of Logic Programming (TPLP).

\end{abstract}
\section{Introduction}
A growing body of research on  scalable data analytics has  brought a renaissance of interest
in Datalog because of its ability to specify declaratively advanced data-intensive applications
that execute efficiently over different systems and architectures, including
massively parallel ones~\cite{seo2013distributed,DBLP:journals/pvldb/ShkapskyZZ13,DBLP:conf/bigdataconf/YangZ14,aref2015design,wang2015asynchronous,yang-iclp2015,bigdatalog,DBLP:journals/vldb/YangSZ17}.
The trends and developments 
that have led to this renaissance can be better
appreciated  if we contrast them with those that
motivated the early research on Datalog 
back in the 80s. The  most obvious difference
is the great importance and pervasiveness of  Big Data  that,
 by enabling intelligent decision making  and solving complex 
problems,  is delivering major benefits to societies and economies.
This is remarkably different from the early work on Datalog in the 80s,
which was motivated  by interest in 
expert system applications that then proved to be only of transient significance.
The main objective of this paper is to present
the significant technological advances that
have made possible for Datalog to exploit 
the opportunities created by
Big Data applications.  One is the newly found
ability to support a larger set of applications by 
extending the declarative framework of Horn clauses to include 
aggregates in recursive rules. The other is the ability
of scaling up Datalog 
applications on Big Data by exploiting  parallel
systems featuring multicore and distributed architectures.
We will  next introduce and discuss the first topic, by summarizing
the recent findings presented in \cite{DBLP:journals/tplp/ZanioloYDSCI17}
and then extending them with new examples of graph algorithms, 
and Knowledge Discovery and Data Mining 
(KDD) applications.
The second topic is briefly discussed in this section;   then it is revisited in Section 5 and  
fully discussed in Sections 6 and 7  on the basis of  results and techniques from  \cite{bigdatalog} and \cite{DBLP:journals/vldb/YangSZ17}.

A common trend in the new generation of
Datalog systems is the usage of aggregates in recursion, 
since they enable
the concise expression and efficient
support of much more powerful algorithms than those expressible by programs that are
stratified w.r.t. negation and aggregates
\cite{seo2013distributed,DBLP:journals/pvldb/ShkapskyZZ13,wang2015asynchronous,bigdatalog}.
As discussed in more detail in the related work section,
extending the declarative semantics of Datalog
to allow aggregates in recursion
represents a difficult problem that had seen much action in the early days 
of Datalog ~\cite{DBLP:conf/slp/KempS91,DBLP:conf/pods/GrecoZG92,ross1992monotonic}.
Those approaches sought to achieve both (i) a formal declarative
semantics for deterministic queries using the basic SQL aggregates, min, max,
count and sum, in recursion
and  (ii) their efficient  implementation by extending techniques  of the early  Datalog 
systems \cite{DBLP:conf/iclp/MorrisUG86,DBLP:journals/debu/ChimentiOKTWZ87,DBLP:conf/vldb/RamakrishnanSS92,DBLP:journals/vldb/VaghaniRKSSLH94,ldl++}.  Unfortunately,
as discussed in the Related Work section, some of those approaches 
had limited generality since they did not deal 
 with all four basic aggregates, while the
proposal presented in \cite{ross1992monotonic} that 
was covering all four basic aggregates using 
different lattices for different aggregates faced other limitations,
including those pointed out by \cite{van1993foundations} that are discussed
in Section~\ref{sec:related}.
These works were followed by  more recent approaches 
that addressed the problem of using more powerful semantics, 
such as answer-set semantics, that require higher levels of computational 
complexity and  thus are a better for higher-complexity problems
than for the very efficient algorithms  needed on Big Data~\cite{DBLP:journals/ai/SimonsNS02,DBLP:journals/tplp/PelovDB07,DBLP:journals/tplp/SonP07,Swift2010,DBLP:journals/ai/FaberPL11,DBLP:journals/corr/GelfondZ14}.

The recent explosion of work on Big Data has also produced a revival
of  interest in Datalog as a parallelizable language
 for  expressing  and  supporting efficiently Big Data Analytics
 ~\cite{seo2013distributed,DBLP:journals/pvldb/ShkapskyZZ13,wang2015asynchronous}. 
 As described from Section 5 onward, the projects discussed in those papers 
 have demonstrated the
 ability of Datalog to provide scalable  support for Big Data applications on both 
 multicore and distributed systems.
Most of the algorithms discussed in those papers are 
graph algorithms or other algorithms that use aggregates in recursion,
whereby a full convergence of formal declarative semantics and
amenability to efficient implementation becomes a critical objective.
%
By supporting graph applications written in Datalog and compiled onto Apache Spark 
with better performance than the same applications written in GraphX (a Spark framework optimized for graph algorithms) and Scala (Spark's native language),
our BigDatalog system \cite{bigdatalog} proved that we have achieved 
this very difficult objective.
 Along with the post-MapReduce  advances demonstrated by  Apache Spark, 
 this success was made possible by
the theoretical developments  presented in \cite{DBLP:journals/tplp/ZanioloYDSCI17}, where the 
concept of \emph{premappability} (\prem ) was introduced for constraints using a unifying semantics
that makes possible  the use of  the aggregates min, max, count and sum 
in  recursive programs. 
Indeed, \prem 
of constraints provides a simple criterion that (i)  the system optimizer
can utilize to push constraints into recursion, and (ii) the user
can  utilize to write  programs using aggregates in recursion,
with the guarantee that they have indeed a formal fixpoint semantics.
 Along with its formal fixpoint semantics, this approach also
 extends the applicability of  traditional  Datalog optimization techniques, 
 to programs that use aggregates  in \label{sec:scalable}rules defining recursive
 predicates.

The rest of this paper is organized as follows. In the next section,
we introduce the problem of supporting aggregates in recursion, then in Section~\ref{sec:graph}
we present how such Datalog extension can be used in practice to implement efficient graph applications. We thus introduce in Section~\ref{sec:analytics} even more advanced
(KDD) analytics such as classification and regression.
Sections~\ref{sec:scalable}, \ref{sec:datalog_spark} and~\ref{sec:datalog_multicore}  introduce our BigDatalog and BigDatalog-MC  systems that support scalable and efficient analytics through distributed and multicore architectures, respectively.
Related work and conclusion presented in Sections \ref{sec:related} and \ref{sec:conclusion}
bring the paper to a closing.
 \newtheorem{theorem}{Theorem}
\def\lfp{{\tt lfp}}
\section{Datalog Extensions: Min and Max 
in Recursive Rules}\label{sec:extrema}

In this section, we first introduce some basics about Datalog before explaining its recent extensions.
A Datalog program  is formally represented as  a  finite  set  of  rules.   
A Datalog rule, in turn, can be represented as 
$h \leftarrow b_1,...,b_n$, 
where
$h$ denotes 
the head of the rule and
$b_1,...,b_n$ represents the corresponding body. 
Technically, $h$ and each $b_i$
are literals assuming the form
$p_i(t_1,...,t_j)$, where $p_i$ is a predicate and each $t_i$ can either be a constant or a variable.
A rule with an empty body is called a \emph{fact}. 
The comma separating literals in a
body  of the rule represents logical  conjunction  (AND). 
Throughout the paper we follow the convention that 
predicate and function names begin with lower case letters, and
variable names begin with upper case letters.

A most significant advance in terms of language and expressive power offered  by our systems \cite{bigdatalog,DBLP:journals/vldb/YangSZ17} is  that they provide a formal semantics and 
efficient implementation  for recursive programs  that use  min, max, count and sum
 in recursion.  We present here an informal summary of  these  advances 
 for which  \cite{DBLP:journals/tplp/ZanioloYDSCI17} provides a formal in-depth coverage.


Consider for instance  Example 1,
where  the goal $\tt is\_min((X,Z), (Dxz))$ in  $\tt r_3$
specifies that we want the min values of $\tt Dxz$ for each
unique pair of values $\tt  (X, Z)$ in $\tt dpath$ defined by rules $\tt r_1$
and $\tt r_2$. 

  \begin{example}[Computing distances between node pairs, and finding their min]
 \label{ex:llimited}
   \vspace{-2ex}
 \bldl
\prule{r_1:dpath(X, Z, Dxz)} {darc(X, Z, Dxz).}
\prule{r_2:dpath(X, Z, Dxz)}{dpath(X, Y,  Dxy) ,  darc(Y, Z,  Dyz),}
\pbody{~~~~~~~~~~~~~~~~~~~~~~~~~~~~~~~~~~~~~~~~~~~~}{ Dxz=Dxy+Dyz.}
\prule{r_3:spath(X, Z, Dxz) } {dpath(X, Z, Dxz),  is\_min((X, Z), (Dxz)).}
\eldl
\end{example}

Thus,  the special notation  $\tt is\_min((X,Z), (Dxz))$ 
tells the compiler that $\tt spath(X, Z, Dxz)$ is a special 
predicate supported by a specialized implementation 
(the  query 
and optimization techniques will be discussed in the following sections). 
Similar observations also hold for $\tt is\_max$.
%
However, the formal  semantics of rules with extrema  constructs is defined using standard (closed-world)  negation,  whereby the semantics of
 $\tt r_3$ is defined by the following two rules\footnote{This rewriting assumes that there is only 
 one $\tt is\_min$ in our program. In the presence of multiple occurrences, we will need to add a subscript to  
keep them distinct.}.
\inv
\bldl
\prule{spath(X, Z, Dxz) } {dpath(X, Z, Dxz),  \neg  lesser(X, Z, Dxz).}
\prule{lesser(X, Z, Dxz)}{dpath(X, Z, Dxz),  dpath(X, Z,  D1), D1 <Dxz.}
\eldl
Expressing  $\tt is\_min$ via negation also reveals
 the  non-monotonic nature of extrema constrains,  whereby this program will be treated as a  stratified program, with a "perfect model" semantics, realized by an iterated-fixpoint  computation~\cite{DBLP:conf/iclp/Przymusinski88}. In this computation, 
 $\tt dpath$ is assigned to a stratum lower than
$\tt spath$ and thus the computation of $\tt dpath$  must 
complete before the computation of $\tt spath$  via $\tt is\_min$ in $\tt r_3$   can begin.
This stratified computation  can be very inefficient or
even non-terminating when the original graph 
of Example 1 contains cycles.
Thus, much research work was  spent on solving this problem,
before the simple  solution described next emerged, and was
used in \bigdatalog to  support graph algorithms with superior performance \cite{bigdatalog}.
This  solution  consists in taking the $\tt is\_min$ 
constraint on $\tt dpath$ in $\tt r_3$ and moving it to the rules 
$\tt r_1$ and $\tt r_2$ defining $\tt dpath$, producing 
the rules in Example 2.
This rewriting   will be called a {\em transfer} 
of constraints\footnote{In the example at hand we  have conveniently used the  same 
names for  corresponding variables  in all our rules. In general  however, the transfer
 also involves a renaming for  the variable(s)  used in specifying the 
 constraint.}.

\begin{example}[Shortest Distances Between Node Pairs]
\label{ex:shortestpath1}
  \vspace{-2ex}
\bldl
\hspace{-1ex}\prule{r'_1:dpath(X, Z, Dxz)} {darc(X, Z, Dxz),   is\_min((X,Z), (Dxz)).}
\hspace{-1ex}\prule{r'_2:dpath(X,Z, Dxz)}{dpath(X, Y,  Dxy) ,  darc(Y, Z,  Dyz),}
\pbody{~~~~~~~~~~~~~~~~~~~~~~~~~~~~~~~~~~~~~~~~~~~~}{ Dxz=Dxy+Dyz,  is\_min((X,Z), (Dxz)).}
\hspace{-1ex}\prule{r'_3:spath(X,Z,  Dxz) } {dpath(X, Z, Dxz).}
\eldl
\end{example}
While, at the syntactic level, this transfer  of constraint is quite simple, at the semantic level,
 it raises the  following 
two  critical questions: (i) does  the program in Example 2 have
a formal semantics, notwithstanding the fact that it uses non-monotonic constructs in recursion,
and (ii)  is it equivalent to the original program,  insofar  as it produces the same  answers for $\tt spath$? 
A positive formal answer  to both  questions was provided in \cite{DBLP:journals/tplp/ZanioloYDSCI17} using 
the notion of {\em premappability} (\prem)  which is summarized next.
\inv
\paragraph*{Premappability (\prem) for Constraints.} Let $T$ denote the Immediate Consequence Operator (ICO)
for the rules defining a  recursive predicate\footnote{The
case of multiple mutually recursive predicates will be
discussed later.}.
Since our rules are positive, the mapping defined by
$T$ has a least-fixpoint  in the lattice of set containment.
Moreover the property that such least-fixpoint is equivalent to the fixpoint iteration
$T^{\uparrow \omega}(\emptyset)$  allows us to turn this declarative semantics into a concrete one.
Now let $\gamma$ be a constraint, such as an extrema constraint like min.
We have the following important definition:

\begin{definition}
The constraint $\gamma$ is said to be \prem
to $T$ when, for every interpretation $I$, we have that: $\gamma(T(I)) = \gamma (T(\gamma (I)))$.
\end{definition}
For convenience of notation, we will also denote by $T_\gamma$ the
composition of   the function $T$  with the function $\gamma$, i.e.,
 $T_\gamma(I)= \gamma(T(I))$, which will be called the {\em constrained
 immediate consequence operator} for $T$  and the rules having $T$  as their ICO.
 Then, \prem holds whenever $T_\gamma (I)=  T_\gamma( \gamma(I))$.
%
%
%
%
%
We will next focus on  cases of practical interest where the transfer of constraints 
under \prem produces optimized programs that are safe and terminating 
(even when the original programs do not terminate). 
Additionally, we prove that the transformation is indeed
equivalence-preserving. Thus we focus on situations where 
$T_\gamma^{\uparrow n}(\emptyset)= T_\gamma^{\uparrow n+1}(\emptyset)$, i.e., 
the fixpoint iteration converges after a finite number of steps $n$. 
The rules defining a  recursive predicate $p$ are those having as head $p$ or predicates that are mutually recursive with $p$. 
Then the following  theorem was proven in ~\cite{DBLP:journals/tplp/ZanioloYDSCI17}:

\inv
\begin{theorem}
\label{thm:1}
In a Datalog program, let  $T$ be the ICO for the positive rules defining a recursive
predicate. 
If the constraint $\gamma$ is \prem to $T$,  and a fixpoint exists such that
$T^{\uparrow n+1}(\emptyset) = T^{\uparrow n}(\emptyset)$ for some integer $n$, then
$\gamma(T^{\uparrow \omega}(\emptyset)) = T_\gamma^{\uparrow n}(\emptyset)$.
\end{theorem}

In~\cite{DBLP:journals/tplp/ZanioloYDSCI17} it was also shown that the fixpoint so 
derived is a minimal fixpoint for the program  produced by the transfer of constraints.
 Thus if a constraint is \prem to the given recursive
 rules, its transfer produces  an optimized program
 having a declarative semantics defined by the minimal fixpoint of
its constrained ICO $(T_\gamma)$ and operational semantics
supported by a terminating fixpoint iteration, with 
 all the theoretical and computational
properties that follow from such semantics.
For instance, \prem for extrema constraints holds  for Example~\ref{ex:llimited}, and since 
directed arcs  in our graph have non-negative lengths,
we conclude  that its optimized version in
Example~\ref{ex:shortestpath1} terminates even if the original graph has cycles.

 For most applications of practical interest, \prem is simple  for   users
 to program with,  and  for the
 system to support\footnote{ In fact, premappability is a very general property that has  been widely used in advanced analytics under
different  names and environments. For instance, the  antimonotonic property of frequent item sets
represents just a particular form of premappability that  will be discussed in Section \ref{sec:analytics}.
Also  with $\tt OP$ denoting  sum or  min  or max,  we have that:
\inv$$ {\tt OP} (\bigcup_{1\leq j \leq K} {\Large S_j}) = {\tt OP} (\bigcup_{1\leq j \leq K} {\tt OP} ({\Large S_j} ))$$
Thus $\tt OP$ is premappable w.r.t. union; this is the  pre-aggregation
 property that  is  commonly used  in distributed processing since it delivers major optimizations \cite{preaggDistComputing}.}.
For instance, to realize that \prem of 
min and max holds for the rules  for our  Example 1,
 the programmer will test \prem by asking  how the mapping established by  rules $\tt r'_1$ and $\tt r'_2$
in Example 2 changes if, in addition to
the post-constraint $\tt is\_min$ that applies to the cost arguments of the head of  rules $\tt r'_1$ and $\tt r'_2$,
we add the goal $\tt is\_min$ 
in the body of our two rules to pre-constrain the values of the cost argument in every $\tt dpath$ goal.
Of course,  \prem is trivially satisfied in  $\tt r'_1$  since this is an exit rule  with no $\tt dpath$ goal,
whereby the rule and its associate mapping remain unchanged.
In rule $\tt r'_2$ the application of the pre-constraint $\tt is\_min((X,Y), Dxy) $ to the values generated by $\tt dpath(X, Y, Dxy)$
does not change the final values returned by this rule because of the arithmetic properties of 
 its interpreted goals $\tt Dxz=Dxy+Dyz$;  in fact, these assure that every $\tt \overline{Dxy}  > Dxy$ can be eliminated since the value  $\tt \overline{Dxz} =\overline{Dxy}+Dyz$  it produces is  higher than $\tt  Dxz$ and  will thus be eliminated by the $\tt is\_min((X,Z), Dxz) $  post-constraint.

This line of reasoning is simple enough for the programmer to understand and for the system to verify. More general conditions
for \prem are given in \cite{DBLP:journals/tplp/ZanioloYDSCI17} using the notions of inflation-preserving and deflation-preserving rules. There, 
we also discuss  the premappability of the lower-bound and the upper-bound constraints  which  are often used
in conjunction with extrema, and  interact with them to determine  \prem and  the termination of the  resulting program.
For instance, to find the maximum distance between nodes in a graph that is free of directed cycles, the programmer  will simply  replace
 $\tt is\_min$  with  $\tt is\_max$ in  Example 1 and  
 Example 2 with the assurance that the second program
 so obtained is the optimized equivalent of the first 
since (i) premappability  holds, and (ii) its computation
terminates in a finite number of steps\footnote{Besides representing a practical requirement in
applications, termination is also required from a theoretical viewpoint since, for
programs such as that of Example 2, a stable model exists if and only if it has a  termination [A. Das and M. Interlandi,
personal communication].}.  However, say that the
programmer wants to add to
the recursive rule of this second program the  condition
$\tt Dxz< Upperbound$ 
either because (i) only results that satisfy this 
inequality are of interest, or (ii) this  
precautionary step is needed to guarantee termination when
fortuitous cycles are created  by 
accidental insertions of wrong data\footnote{For example, 
Bill of Materials (BoM) databases store, for each part in
the assembly, its subparts with their quantities. BoM databases define
acyclic directed graphs; but the risk of some bad data  
can never be ruled out in such databases containing millions of records.}.
However, if the condition $\tt Dxz< Upperbound$
is added as a goal to recursive rule of our  program, its
\prem property is compromised.
To solve this problem the Datalog programmer 
should instead replace $\tt  ~Dyz{=}Dxz{+}Dxy~ $  with the condition:
$$\tt if(Dxy{+}Dyz > Upperbound ~~then~~Dxz{=}Upperbound ~~else~~
  Dxz{=}Dxy{+}Dyz)$$
This condition can be expressed as such in our systems,  or
can be re-expressed using a pair of positive rules in other Datalog systems.
This formulation ensures termination while preserving \prem for max constraints.
Symmetrically, the addition of lower-bound constraints in our Example 2 must be performed in a similar way to avoid
compromising \prem.

Our experience suggests that using 
 the insights gained from these simple  examples, a programmer can master the  use of \prem constraints to
 express significant algorithms  in Datalog, with assurance  they will deliver performance and scalability.  

In the next example, we present a non-linear version of  Example 1, where we 
use the head notation for aggregates that is  supported in our system.
\begin{example}[Shortest Distances Between Node Pairs]
\label{exfloyd}
  \vspace{-2ex}
\bldl
\hspace{-1ex}\prule{(r_4)~~dpath(X, Z, min\langle Dxz\rangle)} {darc(X, Z, Dxz), Dxz>0.}
\hspace{-1ex}\prule{(r_5)~~dpath(X,Z, min\langle Dxz\rangle)}{dpath(X, Y,  Dxy) ,  dpath(Y, Z,  Dyz),Dxz=Dxy{+}Dyz.}
\eldl
\end{example}
The special  head notation, is in fact a short hand for adding
final goal $\tt is\_min((X,Z), (Dxz))$ which still defines the formal semantics of our rules. Therefore,
\prem for $\tt r_5$ is determined by adding the  pre-constraints  $\tt is\_min((X,Y), (Dxy))$
and  $\tt is\_min((Y,Z), (Dxy))$ respectively  after the first and the second goal and  asking
if  these changes affect the final values that survive the post-constraint in the head of the rule. Here again, the 
  values  $\tt \overline{Dxy} >Dxy$  and values  $\tt \overline{Dyz} >Dyz$  can be eliminated without changing the
head results once  the post constraint is applied.

\inv\subsection{From Monotonic Count  to Regular COUNT and SUM}
\label{sec:count}
At the core of the approach proposed in \cite{mazuran2012extending}
there is the observation that the cumulative version of standard count is
monotonic in the lattice of set containment. Thus the authors introduced  
 $\tt mcount$  as the 
aggregate function that returns all natural numbers up to the cardinality of the set.
The use of $\tt mcount$ in actual applications
is illustrated by the following example that uses the  motonic count
$\tt mcount$ in the head of rules to express an application similar to the one 
proposed in \cite{ross1992monotonic}.

\begin{example}[Join the party once you see that three of your friends have joined]
\label{ex:attend1}
The organizer of the party will attend, while other people
will attend if the number of their friends attending 
is  greater or equal to 3, i.e., ${\tt Nfx \geq 3}$.
\bldl
\prule{r_6: attend(X)}{organizer(X).}
\prule{r_7: attend(X)}{cntfriends(X, Nfx), Nfx \geq 3.}
\prule{r_8: cntfriends(Y, mcount\langle X\rangle)}{attend(X), friend(Y,X).}
\prule{r_9: finalcnt(Y, max\langle N\rangle)}{cntfriends(Y, N).}
\eldl
\inv\inv
\end{example}


As described in \cite{mazuran2012extending}, the formal semantics of  $\tt mcount$
can be reduced to the formal semantics of Horn Clauses.
Thus,  $\tt mcount$ is a monotonic aggregate function and as such is fully compatible with standard  semantics of Datalog 
and its optimization techniques, including the transfer of extrema, discussed in the previous section.
In terms of operational semantics however, $\tt mcount$ will  enumerate new
friends one at the time and could  be somewhat slow.  An obvious alternative
consists in premapping the max value to $\tt mcount$ since the combination 
of $\tt mcount$ and $\tt max$ defines the traditional count. Then in  the fixpoint
computation, the new count value will be upgraded to the new max, rather than the
succession of $+1$ upgrades computed  by $\tt mcount$. 
Thus the rules $\tt r_8, r_9$ can be substituted with $\tt r'_8, r'_9$ respectively as follows:
\inv
\bldl
\prule{r'_8: cntfriends(Y, count\langle X\rangle)}{attend(X), friend(Y,X).}
\prule{r'_9: finalcnt(Y, N)}{cntfriends(Y, N).}
\eldl
The question of 
whether $\tt max$ is \prem to our rules can be formulated by  assuming
that we apply a vector of constraints one for each mutually recursive predicate.
Thus, in the Example \ref{ex:attend1}, we will apply the max constraint to $\tt cntfriends$ and a null constraint,
that we will call $\tt nofilter$, to $\tt attend$.  Now, the addition
of $\tt nofilter(X)$ does not change the mapping defined by $\tt r_8$,
and the addition of $\tt is\_max(X, Nfx)$ does not change the mapping 
defined by $\tt r_7$ since the condition  $\tt Nfx  \geq 3$ is satisfied for some 
$\tt Nfx$ value iff  it is satisfied by the  max of these values. Thus 
\prem is satisfied and $\tt mcount$ in $\tt r_8$ can be
replaced by the regular $\tt count$.

\paragraph{From Monotonic SUM to  SUM.} 
The notion of monotonic sum, i.e.,  $\tt msum$, for positive numbers
introduced in \cite{mazuran2012extending}  
uses the fact that its semantics   can be easily reduced to that of $\tt mcount$,
as  illustrated by the example below that computes the total number of  each part 
available in a city
by adding up the quantities held in each store  of that city:

\cldl
\prule {\hspace{-3ex}pCnt\_InCity(Pno, City, sum\langle Qty, Store \rangle)}   {pqs(Pno, Store, Qty),  cs(Store, City).}
\eldl

Here, the sum  is computed by adding up the
$\tt Qty$ values,  but  the presence of $\tt Store$ makes sure 
that  all the repeated occurrences of the same $\tt Qty$ are considered in the addition,
rather than being ignored as a set semantics would imply.
The results returned by this rule are the same as those returned by the following  rule
where $\tt posint$ simply enumerates the positive integers up to  $\tt Qty$:

\bldl
\prule {partCnt\_InCity(Pno, City, count \langle Eq, Store\rangle)}   {}
\pbody{\hspace{3cm}} {pqs(Pno, Store, Qty),  cs(Store, City),posint(Qty, Eq).}
\eldl

Then consider the following example where we want to count the distinct paths
connecting any pair of node in the graph:

\cldl
\prule{cpath(X, X, 1)} {arc(X, \_).}
\prule{cpath(X, Z, sum \langle Cxy, Y \rangle)} { cpath(X, Y, Cxy), arc(Y, Z).}
\eldl
Then the semantics of our program is defined by its equivalent rewriting 
\begin{example} [Sum of positive numbers expressed via count. ]
\inv\bldl
\prule{cpath(X, Y, 1)} {edge(X, Y).}
\prule{cpath(X, Z,  count \langle Y, Ixy \rangle)}  {cpath(X, Y, Cxy), edge(Y, Z), posint(Cxy, Ixy).}
 \eldl                           
\end{example}
Thus, whenever a sum aggregate is used, the  programmer and the compiler will
determine its correctness, by (i) replacing  sum with msum,  (ii) replacing  msum
by mcount via the $\tt posint$ expansion, and (iii) checking that the max aggregate
is \prem in the program so rewritten.  Of course, once  this check succeeds, the
actual implementation  uses  the sum aggregate directly, rather than its
equivalent, due to the inefficient expansion of the count aggregator. 
While, in this example, we have used positive integers for cost arguments, the sum of positive floating point 
numbers can also be handled in the same fashion \cite{mazuran2012extending}.

\section{In-Database Graph Applications}\label{sec:graph}

The use of aggregates in recursion has allowed to express efficiently
a wide spectrum of applications that were very difficult to express
and support in traditional  Datalog. Several graph and mixed graph-relation applications
were described in \cite{bigdatalog} and \cite{yang2017declarative}. Other 
 applications\footnote{Programs available at \url{http://wis.cs.ucla.edu/deals/}} 
include, the Viterbi algorithm for Hidden Markov models, Connected Components by Label Propagation,
Temporal  Coalescing of  closed periods, the People you Know, the Multi-level Marketing Network Bonus Calculation, and several Bill-of-Materials queries such as parts, costs,
and days required in an assembly. 
Two new graph  applications that we have recently developed are given next,
and advanced analytics and  data mining applications are discussed in the next section.

\begin{example}[Diameter Estimation]\label{rule:diamest}

Many graph applications, particularly those appearing in the social networks setting, need to make an estimation about the diameter of its underlying network in order to complete several critical graph mining tasks like tracking evolving graphs over time \cite{DiamEst}. The traditional definition of the diameter as the farthest distance between two connected nodes is often susceptible to outliers. Hence, we compute the \emph{effective diameter}~ \cite{DiamEst} which is defined as follows:
the effective diameter $d$ of a graph
$G$
is formally defined as the
minimum number of hops in which 90\% of all connected pairs of nodes can reach each
other. This measure is tightly related to closeness centrality and in fact is widely adopted in many network mining tasks \cite{diamEstReference}. The following Datalog program shows how effective diameter can be estimated using  aggregates in recursion.

\inv\begin{displaymath}
\begin{aligned}
& r_{\ref{rule:diamest}.1}: {\tt hops(X, Y, H)} \arrow  {\tt arc(X, Y)}, {\tt H = 1}.\\
& r_{\ref{rule:diamest}.2}: {\tt hops(X, Y, \minv{C})} \arrow  {\tt hops(X, Z, C1)},  {\tt hops(Z, Y, C2)}, {\tt C = C1 + C2}.\\
& r_{\ref{rule:diamest}.3}: {\tt minhops(X, Y, C)} \arrow  {\tt hops(X, Y, C)}.\\
& r_{\ref{rule:diamest}.4}: {\tt totalpairs( \cntv{X} )} \arrow  {\tt minhops(X,\_, \_)}.\\
& r_{\ref{rule:diamest}.5}: {\tt cumulhops(C, \cntv{(X, Y)})} \arrow {\tt minhops(X, Y, C)}.\\
& r_{\ref{rule:diamest}.6}: {\tt cumulhops(H2, \sumv{(H1, C)})} \arrow {\tt cumulhops(H1, C1)},
{\tt cumulhops(H2, C2)}, \\
& \hspace{17.3em} {\tt H1 < H2}, {\tt C = C1 + C2}.\\ 
& r_{\ref{rule:diamest}.7}: {\tt effdiameter(\minv{H})} \arrow {\tt cumulhops(H, C)}, {\tt totalpairs(N)}, {\tt C/N \geq 0.9}.
\end{aligned}
\end{displaymath}

\vspace{5pt}

Rules $r_{\ref{rule:diamest}.1}$-$r_{\ref{rule:diamest}.3}$ find the minimum number of hops for each connected pair of vertices whereas rules $r_{\ref{rule:diamest}.5}$-$r_{\ref{rule:diamest}.6}$ compute the cumulative distribution of hops recursively using the fact that any pair of connected vertices covered within $H1$ hops is also covered in $H2$ hops ($H1<H2$). The final rule $r_{\ref{rule:diamest}.7}$ extracts the effective diameter as per its definition \cite{DiamEst}.
\inv
\end{example}

\begin{example}[$k$-Cores Determination]\label{rule:kcore}

A $k$-core of a graph $G$ is a maximal connected subgraph of $G$ in which all vertices have degree of at least $k$. $k$-core computation \cite{KCore} is critical in many graph applications to understand the clustering structure of the networks and is frequently used in bioinformatics and in many network visualization tools \cite{kcoreApp}. The following Datalog program computes all the $k$-cores of a graph for an input $k$. Using 
aggregates in recursion in the following computation we determine all the connected components of the corresponding subgraph with degree $k$ or more. 

\vspace{-6pt}\begin{displaymath}
\begin{aligned}
& r_{\ref{rule:kcore}.1}: {\tt degree(X,\cntv{Y})} \arrow {\tt arc(X, Y)}. \\
& r_{\ref{rule:kcore}.2}: {\tt validArc(X, Y)} \arrow {\tt arc(X, Y)}, {\tt degree(X, D1)}, {\tt D1 \geq k}, \\ 
& \hspace{15.3em} {\tt degree(Y, D2)}, {\tt D2 \geq k}. \\
& r_{\ref{rule:kcore}.3}: {\tt connComp(A,A)} \arrow {\tt validArc(A, \_)}. \\
& r_{\ref{rule:kcore}.4}: {\tt connComp(C, \minv{B})} \arrow {\tt connComp(A, B)}, {\tt validArc(A, C)}. \\
& r_{\ref{rule:kcore}.5}: {\tt kCores(A,B)} \arrow {\tt connComp(A, B)}. \\
\end{aligned}
\end{displaymath}
\vspace{2pt}

Example \ref{rule:kcore}
determines $k$-cores by determining all the connected components ($r_{\ref{rule:kcore}.3}$, $r_{\ref{rule:kcore}.4}$), considering only vertices with degree $k$ or more ($r_{\ref{rule:kcore}.1}$, $r_{\ref{rule:kcore}.2}$).
The lowest vertex id is selected as the connected component id among the $k$-cores. 

\end{example}

\section{Advanced Analytics}\label{sec:analytics}

The application area of ever-growing importance, advanced analytics, 
 encompass applications using standard OLAP to complex data mining and machine learning queries like frequent itemset mining \cite{agrawal1994fast}, building classification models, etc. This new generation of 
advanced analytics is extremely useful in extracting meaningful and rich insights from data \cite{agrawal1994fast}. 
However, these advanced analytics have  
created  major challenges to database researchers \cite{agrawal1994fast} and 
the Datalog community \cite{ldl++,DBLP:conf/jelia/GiannottiM02,DBLP:journals/tkde/GiannottiMT04,DBLP:journals/debu/BorkarBCRPCWR12}. 
The major success that BigDatalog has achieved on graph algorithms
suggests that we should revisit this hard problem   
and look beyond the initial applications discussed in \cite{reviewer2-Tsur}   
by leveraging on
the new opportunities created by the use of aggregates in recursion.
We  next describe
briefly the approach we have taken and the results obtained so far.



\inv\paragraph*{Verticalized Representation.}
Firstly, we need to specify  algorithms that can support advanced analytics
on tables with arbitrary number of columns. A simple way to achieve
this genericity is to use verticalized representations for tables. For instance,
consider the excerpt from the well-known {\em PlayTennis} example from \cite{mitchell1997machine}, shown in \ctab\ref{tab:train}. The corresponding verticalized view is presented in \ctab\ref{tab:vtrain}, where each row contains the original tuple ID, a column number, and the value of the corresponding column, respectively.  
The verticalization of  a table with $n$ columns (excluding the ID column) can be easily  expressed by  $n$ rules, however a special ``$\tt @$'' construct is provided in our language to expedite this task.
The use of this special construct is demonstrated by the rule below, which converts \ctab\ref{tab:train} into the verticalized view of \ctab\ref{tab:vtrain}.
\inv
\bldl
\prule{vtrain(ID, Col, Val)}{train(ID, Val@Col).}
\eldl

Given a vertical representation,
a simple data mining algorithm such  as Naive Bayesian Classifiers \cite{NBCPaper} 
can  be  expressed by simple
non-recursive rules\footnote{http://wis.cs.ucla.edu/deals/tutorial/nbc.php}.
However a more advanced compact representation is needed  to  support complex tasks efficiently,  
as outlined next.  

\begin{minipage}[t]{0.6\linewidth}
\centering
\captionof{table}{Training examples for the PlayTennis table.}\label{tab:train}
\vspace{-2ex}
{\small
 \begin{tabular}{cccccc}
 \hline
 \multirow{2}{*}{\bf ID} & \multirow{2}{*}{\bf Outlook} & {\bf Tempe} & \multirow{2}{*}{\bf Humidity} & \multirow{2}{*}{\bf Wind} & {\bf Play} \\
 & & {\bf rature} & & & {\bf Tennis} \\
 & (1) & (2) & (3) & (4) & (5) \\
 \hline
 \hline
 1	 & overcast	 & cool		 & normal	 & strong    & yes \\	 
 2	 & overcast	 & hot		 & high		 & weak	     & yes \\
 3	 & overcast	 & hot		 & normal	 & weak      & yes \\
 4	 & overcast	 & mild		 & high		 & strong    & yes \\	
5	 & rain		 & mild		 & high		 & weak	     & yes \\
6	 & rain		 & cool		 & normal	 & weak      & yes \\
7	 & rain		 & cool		 & normal	 & strong    & no \\	
8	 & rain		 & mild		 & high		 & strong    & no \\
9	 & rain		 & mild		 & normal	 & weak      & yes \\
10	 & sunny     & hot		 & high      & weak      & no \\
$\ldots$	 & $\ldots$		 & $\ldots$		 & $\ldots$		 & $\ldots$	     & $\ldots$ \\
\end{tabular}
}
\end{minipage}
\hspace{7ex}
\begin{minipage}[t]{0.25\linewidth}
\centering
\captionsetup{justification=centering}
\captionof{table}{Vertical view of the tuples in \ctab\ref{tab:train}.}\label{tab:vtrain}
\vspace{-2ex}
{\small
 \begin{tabular}{ccc}
 \hline
 {\bf ID} & {\bf Col} & {\bf Val} \\
 \hline \hline
 1	 & 1     & overcast \\	
 1	 & 2  	 & cool \\	
 1	 & 3     & normal	\\	
 1	 & 4     & strong \\
 1   & 5     & yes \\
 2	 & 1     & overcast \\
 2	 & 2     & hot \\	
 2	 & 3     & high \\	
 2	 & 4     & weak \\
 2	 & 5     & yes \\		
 $\ldots$ & $\ldots$ & $\ldots$ \\
 \end{tabular}
}
\end{minipage}
\vspace{1em}



\inv\paragraph*{Rollup Prefix Table.} To support efficiently more complex algorithms, such as frequent itemset mining \cite{agrawal1994fast} and decision tree construction~\cite{QuinlanDecTree}, we  use an intuitive  prefix-tree like representation that is
basically a compact representation of 
the {\sc SQL-2003 count rollup}\footnote{https://technet.microsoft.com/en-us/library/bb522495(v=sql.105).aspx} aggregate. For instance, the count rollup on \ctab\ref{tab:train} yields the output of \ctab\ref{tab:table3}, where
we limit the output to the first 14 lines.

\begin{table*}[htbp]
\begin{minipage}[t]{0.62\linewidth}
\caption{{The {\sc SQL-2003 count rollup} on \ctab\ref{tab:train}}.}\label{tab:table3}
\vspace{-2ex}
{\footnotesize 
 \begin{tabular}{ccccccc}
 \hline
 {\sc RID} & {\bf Outlook} & {\bf Temp.} & {\bf Humidity} & {\bf Wind} & {\bf Play} & {\emph{count}} \\
 & (1) & (2) & (3) & (4) & (5) & \\
 \hline \hline
 1	 & \textcolor{red}{null}		 & null		 & null	 & null	 & null & 14 \\
 2	 & \textcolor{red}{overcast}	 & null		 & null	 & null	 & null & 4 \\
 3	 & overcast	 & \textcolor{red}{cool}		 & null	 & null	 & null & 1 \\
4	 & overcast	 & cool		 & \textcolor{red}{normal}	 & null	 & null & 1 \\
 5	 & overcast	 & cool		 & normal	 &  \textcolor{red}{strong}	& null & 1 \\
 6	 & overcast	 & cool		 & normal	 &  strong	& \textcolor{red}{yes} & 1 \\
 7	 & overcast	 & \textcolor{red}{hot}		 & null	 & null	 & null & 2 \\
 8	 & overcast	 & hot		 & \textcolor{red}{high}       &null & null & 1 \\
 9	 & overcast	 & hot		 & high       & \textcolor{red}{weak} & null& 1 \\
 10	 & overcast	 & hot		 & high       &weak & \textcolor{red}{yes} & 1 \\
 11	 & overcast	 & hot		 & \textcolor{red}{normal}       & null  & null & 1 \\
12	 & overcast	 & hot 		 & normal       & \textcolor{red}{weak}  & null & 1 \\
13	 & overcast	 & hot 		 & normal       & weak  & \textcolor{red}{yes} & 1 \\
14	 & overcast	 & \textcolor{red}{mild} & null       & null  & null & 1 \\
$\ldots$ & $\ldots$ & $\ldots$ & $\ldots$ & $\ldots$ & $\ldots$ & $\ldots$
\end{tabular}
}
\end{minipage}
\hspace{10pt}
\begin{minipage}[t]{0.34\linewidth}
\centering
\caption{{A \emph{rollup prefix table}.}}\label{tab:table4a}
\vspace{-2ex}
{\footnotesize 
 \begin{tabular}{ccccc}
 \hline
 {\bf ID} & {\bf Col} & {\bf Val} & {\emph{count}} & {\bf PID}\\ \\
 \hline\hline
 1  & 1 & \textcolor{red}{null}      & 14 & 1\\
 2  & 1 & \textcolor{red}{overcast}  & 4 & 1\\
 3  & 2 & \textcolor{red}{cool}      & 1 & 2\\
 4  & 3 & \textcolor{red}{normal}	& 1 & 3\\
 5  & 4 & \textcolor{red}{strong}	& 1 & 4\\
 6  & 5 & \textcolor{red}{yes}       & 1 & 5\\
 7  & 2 & \textcolor{red}{hot}       & 2 & 2\\
 8  & 3 & \textcolor{red}{high}      & 1 & 7\\
 9  & 4 & \textcolor{red}{weak}      & 1 & 8\\
 10  & 5 & \textcolor{red}{yes}      & 1 & 9\\
 11  & 3 & \textcolor{red}{normal}     & 1 & 7 \\
   12 & 4 & \textcolor{red}{weak}      & 1 & 11\\
 13 & 5 & \textcolor{red}{yes}    & 1 & 12\\
 14 & 2 & \textcolor{red}{mild}       & 1 & 2\\
 $\ldots$ & $\ldots$ & $\ldots$ & $\ldots$ & $\ldots$
\end{tabular}
}
\end{minipage}
\end{table*}

%
%
%

Interestingly, the output of {\sc rollup} contains many redundant \verb|null| values and
only the items in the main diagonal hold new information  (highlighted in red).
In fact, the items to left of the diagonal are repeating the previous values (i.e. sharing the prefix), whereas those to right are \verb|nulls|.
With this observation, we can compact \ctab\ref{tab:table3} to a more logically concise representation shown in \ctab\ref{tab:table4a}, where the first four columns contain the same information as an item in the main diagonal does, 
whereas the last column (PID) specifies the ID of the parent tuple from where we can find the value of the previous column. We refer to this condensed representation as a {\em prefix table} since it is
in fact a  table representation of the well-known prefix tree data structure. 
In this particular case, we have a {\em rollup prefix table} for count, and similar representations can be used for other aggregates.


An easier way to understand and visualize \ctab\ref{tab:table4a} is through
the logically equivalent and more user intuitive representation, {\em compact rollups}, shown in \ctab\ref{tab:table5} (equivalent values in \ctab\ref{tab:table3}, \ctab\ref{tab:table4a} and \ctab\ref{tab:table5} are marked in red). In this representation, each item $e$ that is not under the ID column and is not empty, represents a tuple in the rollup prefix table, where the values for ID, Col, Val, count columns are the tuple ID of $e$, $e$'s column number, $e$'s value, and the number associated with $e$'s value, respectively.  
Thus \ctab\ref{tab:table4a} captures in a verticalized form the information that in a horizontal form is displayed by  \ctab\ref{tab:table5}. In turn, this is a significantly compressed version of \ctab\ref{tab:table3}.

\begin{table}[t]
\centering
\caption{A compact rollup for the example in \ctab\ref{tab:train}.}\label{tab:table5}
\vspace{-2ex}
{\small
 \begin{tabular}{cc|cc|cc|cc|cc}
 \hline
 {\bf Outlook} & {C1} & {\bf Temperature} & {C2} & {\bf Humidity} & {C3} & {\bf Wind} & {C4} & {\bf Play} & {C5} \\
 \hline \hline
    \textcolor{red}{overcast} & \textbf{4} & \textcolor{red}{cool} & \textbf{1} & \textcolor{red}{normal} & \textbf{1} & \textcolor{red}{strong} & \textbf{1} & \textcolor{red}{yes} & \textbf{1}\\
     &  & \textcolor{red}{hot} & \textbf{2} & \textcolor{red}{high} & \textbf{1} & \textcolor{red}{weak} & \textbf{1} & \textcolor{red}{yes} & \textbf{1}\\
    &   &   &   & \textcolor{red}{normal} & \textbf{1} & \textcolor{red}{weak} & \textbf{1} & \textcolor{red}{yes} & \textbf{1} \\
    &   & \textcolor{red}{mild} & \textbf{1} & high & 1 & strong & 1 & yes & 1\\		
\hline
\end{tabular}
}
\vspace{-2ex}
\end{table}


These rollups are simple to generate from our verticalized representation and they provide a good basis for programming other analytics~\cite{crucialPatterns,yang2017declarative}.
We illustrate the construction of the rollup prefix table from the corresponding verticalized representation, using the rules described in the next example which exploits aggregates in recursion.

\begin{example}[From a verticalized view $\tt vtrain$ to a rollup prefix table]\label{rule:pt1}
Given two rows ${\tt T1}$ and ${\tt T2}$, we say that the row ${\tt T1}$ can {\em represent} the row ${\tt T2}$ (or ${\tt T1}$ can represent ${\tt T2}$ for short) for the first ${\tt C}$ columns if both rows are identical in the first ${\tt C}$ columns (i.e. their prefixes are the same). ${\tt repr}$ is a recursive relation that represents ${\tt vtrain}$ in a different format, where each tuple ${\tt (T, C, V)}$ in ${\tt vtrain}$ is augmented with one more column ${\tt T1}$ indicating that ${\tt T1}$ can represent ${\tt T}$ in first ${\tt C - 1}$ columns, i.e., the parent ID of the current row is ${\tt T1}$. Then a prefix table ${\tt rupt}$ is constructed (without the node count) on top of ${\tt repr}$ in $\tt r_1$, 
where among all the rows with the same parent ID ${\tt Ta}$, and the same value ${\tt V}$ in column ${\tt C}$, the one with the minimal ID ${\tt T}$ is selected as a {\em representative} by the 
aggregate $\min$. 

\inv\begin{displaymath}
\begin{aligned}
& r_{\ref{rule:pt1}.1}: {\tt repr(T1, C, V, T)} \arrow
{\tt vtrain(T, C, V)}, {\tt C = 1}, {\tt T1 = 1}. \\
& r_{\ref{rule:pt1}.2}: {\tt rupt(\minv{T}, C, V, Ta)} \arrow
{\tt repr(Ta, C, V, T)}. \\
& r_{\ref{rule:pt1}.3}: {\tt repr(T1, C, V, T)} \arrow
{\tt vtrain(T, C, V)}, {\tt C1 = C - 1}, {\tt repr(Ta, C1, V1, T)}, \\ 
& \hspace{11.3em} {\tt rupt(T1, C1, V1, Ta)}. \\
\end{aligned}
\end{displaymath}


Assuming we want the rollup prefix table for count (\ctab\ref{tab:table4a}), we can extract the node count 
using the aggregate $\tt count \langle TID \rangle$ 
outside the recursion to derive the final 
table \emph{myrupt} as follows.
\inv\begin{displaymath}
\begin{aligned}
& r_{\ref{rule:pt1}.4}: {\tt myrupt(T, C, V, \cntv{TID}, Ta)} \arrow {\tt rupt(T, C, V, Ta)}, {\tt repr(Ta, C, V, TID)}. \\
\end{aligned}
\end{displaymath}

In this example, the aggregate \texttt{count} could  be transferred into recursion, but it would not save any computation time. 
However, if we want to further use anti-monotonic constraints like (\texttt{COUNT} $\geq k$)\footnote{Often used in iceberg queries \cite{Fang98computingiceberg} and frequent itemset mining \cite{agrawal1994fast}.} to prune many of the nodes from the prefix tree, then pushing count into recursion is a computationally efficient choice. Moreover in the example, since the generation of the counts connected with the rollup prefix table is top-down, such lower-bound anti-monotonic constraints are \prem. The popular Apriori constraint \cite{agrawal1994fast} used in frequent itemset mining is a well-known example that exploits \prem. 
It is also important to point out that 
generation of other aggregates like \texttt{max} and \texttt{min} on the rollup prefix table
can be performed efficiently in a bottom-up manner 
when we have \prem constraints. In fact, the rollup computation can be stopped for \emph{(i)} \texttt{max} when it fails the lower bound constraint and \emph{(ii)} \texttt{min} when it fails the upper bound constraint.

\end{example}

\begin{example}[Computing length of the longest maximal pattern from a rollup prefix table]\label{rule:pt2}

Many data mining applications extract condensed representations like maximal patterns \cite{KDD20YrSurvey} from rollup prefix-tree like structures (e.g. Frequent-Pattern Tree or FP-tree \cite{han2000mining}). More recently, interesting mining applications have been developed, which depend on computing the length of the longest maximal pattern from a FP-tree\footnote{A FP-tree is logically equivalent to a rollup prefix table.} \cite{LongestMaxPatt}.
The following Datalog program performs this task by using 
aggregates in recursion on the rollup prefix table for count \texttt{myrupt}. 

\inv\begin{displaymath}
\begin{aligned}
& r_{\ref{rule:pt2}.1}: {\tt items(C, V, \sumv{Cnt})} \arrow {\tt myrupt(\_, C, V, Cnt, \_)}. \\
& r_{\ref{rule:pt2}.2}: {\tt freqItems(C, V)} \arrow {\tt items(C, V, Cnt)}, {\tt Cnt \geq k}. \\
& r_{\ref{rule:pt2}.3}: {\tt len(T, 0)} \arrow {\tt myrupt(T, C, V, \_, \_)}, {\tt \neg myrupt(\_, \_, \_, \_, T)}, {\tt \neg freqItems(C, V)}. \\
& r_{\ref{rule:pt2}.4}: {\tt len(T, 1)} \arrow {\tt myrupt(T, C, V, \_, \_)}, {\tt \neg myrupt(\_, \_, \_, \_, T)}, {\tt freqItems(C, V)}. \\
& r_{\ref{rule:pt2}.5}: {\tt len(T, \maxv{L})} \arrow 
{\tt len(TC, L1)}, {\tt myrupt(TC, \_, \_, \_, T)}, {\tt myrupt(T, C, V, \_, \_)}, \\
& \hspace{10.8em} {\tt \neg freqItems(C, V)}, {\tt L = L1}. \\
& r_{\ref{rule:pt2}.6}: {\tt len(T, \maxv{L})} \arrow 
{\tt len(TC, L1)}, {\tt myrupt(TC, \_, \_, \_, T)}, {\tt myrupt(T, C, V, \_, \_)}, \\
& \hspace{10.8em} {\tt freqItems(C, V)}, {\tt L = L1 + 1}. \\
& r_{\ref{rule:pt2}.7}: {\tt longest(\maxv{L})} \arrow 
{\tt len(\_, L)}. \\
\end{aligned}
\end{displaymath}

Rules $r_{\ref{rule:pt2}.1},r_{\ref{rule:pt2}.2}$ first compute singleton frequent items (i.e. items occurring above a threshold $k$) which are identified by the column number and its value. Since the longest maximal pattern occurs along the path from the leaf to the root of the prefix tree, rules $r_{\ref{rule:pt2}.5}, r_{\ref{rule:pt2}.6}$ recursively compute the maximum pattern length at each node 
from its descendants in a bottom-up manner from the leaves (selected by rules $r_{\ref{rule:pt2}.3}, r_{\ref{rule:pt2}.4}$) to the root. 
In addition to this, several other advanced analytics like iceberg queries \cite{Fang98computingiceberg}, frequent itemset mining \cite{agrawal1994fast} and decision tree construction \cite{QuinlanDecTree} can be performed efficiently exploiting this rollup prefix table, as it has been pointed out in detail in  
\cite{yang2017declarative}.

\end{example}

\section{Performance and Scalability with Multicore and Distributed Processing}\label{sec:scalable}
Multicore and distributed systems  were developed along different technology paths to provide two 
successful competing  solutions to the problem of achieving  scalability via parallelism. 
For a long time, Moore's Law meant that  programmers could virtually double the speed of their software  by  updating the hardware. But starting in 2005, circa, it became impossible  to double  transistor densities every two years.
Since then  therefore,  computer manufacturers 
exploring alternative ways to increase performance
developed  the very successful computer line of  \emph{multicore processing} systems.
Around the same time (2004-2005),
the distributed processing approach to scalability  used in cluster computing
was developed   by big users of big data. This approach was first developed by  database vendors, such as Teradata, and then popularized 
by  web companies such as Google and Yahoo!, 
since they realized that \emph{distributed processing} among their large clusters of shared-nothing computers provide 
an effective method to process their  large and fast-growing data sets. 
 The growing popularity of the distributed processing approach has been both the cause and the  result
 of  better programming support for  parallel applications: for instance in MapReduce~\cite{mapreduce} users have only to provide a \emph{map} and \emph{reduce} program, while the system takes care of low level details such as data communication, process scheduling and fault tolerance.
 Finally, a major advance in  usability  was delivered by
 Apache Spark  which provides higher level APIs 
 \comment{To aid the developer  in programming distributed computation, the MapReduce framework~\cite{mapreduce} was introduced by Google in 2004, and its open source version, Hadoop, was later developed by Yahoo! in 2008. In MapReduce, data is partitioned and stored in a distributed file system (e.g., GFS or HDFS). Computation is then divided into three steps: a \emph{map step} in which a user-defined map function is applied, in parallel, over the distributed data to produced a set of key/value pairs. The \emph{shuffle step} is then triggered once all the map process have concluded, and map outputs are redistributed in the cluster so that records with the same key are co-located on the same machine. Finally, the \emph{reduce step} is applied in parallel via a user-defined function over the total set of values per key to produce a final value  for each key. The MapReduce framework takes care of low level details such as data communication, process scheduling and fault tolerancy. 
 Finally a major advance of  usability  was delivered by
 Apache Spark  which provides higher level APIs for job specification and 
 managing parallelism  as briefly discussed in the next section.
 For now,  we would like to 
 point out that
 }
  that have made possible the development of  languages and systems supporting
 critical application areas, such as Database applications written in SQL, graph applications using GraphX, and
 data mining  application suites.  But Datalog can go beyond these advances by (i) providing  unified declarative
 framework to support different  applications, and (ii) achieving portability over different parallel systems.
 The significance of point (i) is underscored by the fact that BigDatalog was able to outperform GraphX on
 graph applications \cite{bigdatalog}, and the importance of point (ii) is demonstrated by the fact  that while 
 our Datalog applications will execute efficiently on both Apache Spark and multicore systems, the  porting of
 parallel applications from the former platform to the latter can be quite challenging even for an expert 
 programmer~\footnote{Non trivial optimization techniques such as those presented in Section~\ref{sec:datalog_multicore}, could be
 necessary in general.}.
 
%

In the rest of the paper, we discuss the techniques used in ensuring that  declarative programs expressed in Datalog have performance comparable to hand-written parallel programs on specialized domain-specific languages  running on clusters of distributed shared-nothing computers~\cite{bigdatalog} and a multicore machine~\cite{DBLP:journals/vldb/YangSZ17}. 
\comment{
\subsection{Multicore Architecture}
Multicore machines are composed by one or more processors, where each of them contains several cores on a chip~\cite{DBLP:journals/corr/abs-1110-3535}. Each core is composed of computation units and caches, while the main memory is commonly shared. While the individual cores do not necessarily run as fast as the highest performing single-core processors, they are able to obtain high performance by handling more tasks in parallel.
In this paper we will consider multicore processors implemented as a group of \emph{homogeneous cores}, where the same computation logic is applied in a divide-and-conquer way over a partition of the input dataset. 

Unfortunately, single-core applications do not get faster automatically on a multicore architecture with the increase of cores. For this reason, programmers are forced to write 
specific parallel logic to exploit the performance of multicore architectures. In Section~\ref{sec:datalog_multicore} we will see how Datalog can be used to ease the development of high-performance data-intensive parallel programs.

\subsection{Apache Spark}
 \label{sec:sparkbackground}
 
Apache Spark \cite{zaharia2012resilient} is attracting a great deal of interest as a general platform for large-scale analytics, particularly because of its support for in-memory iterative analytics.  Spark enhances the MapReduce programming model by providing a language-integrated Scala API that enables the expression of programs as \emph{dataflows of second order transformations} (e.g., \texttt{map}, \texttt{filter}) on \emph{Resilient Distributed Datasets} (RDD) \cite{zaharia2012resilient}. An RDD is a distributed shared memory abstraction representing a partitioned dataset. RDDs are immutable, and transformations are coarse-grained and thus apply to all items in the RDD to produce a new RDD.  RDDs can be explicitly cached by the programmer in memory or on disk at workers.  RDDs provide fault tolerance by recomputing the sequence of transformations for the missing partition(s).

Once a Spark job is submitted, the scheduler groups transformations that can be pipelined into a single \emph{stage}. Stages are executed \emph{synchronously} in a topological order: a stage will not be scheduled until all stages it is dependent upon have finished  successfully.  Similar to MapReduce,  Spark \textit{shuffles} between stages to \textit{repartition} outputs among the nodes of the cluster. 
Spark has libraries for structured data processing (Spark SQL), stream processing (Spark Streaming), machine learning (MLlib), and graph processing (GraphX).  Section \ref{sec:datalog_spark} discusses in detail how \bigdatalogspark extends Spark for efficient Datalog evaluation.
}

\section{Datalog on Apache Spark}
\label{sec:datalog_spark}

In this section we provide a summary of \bigdatalogspark \cite{bigdatalog}, a full Datalog language implementation on Apache Spark. \bigdatalogspark\ supports relational algebra, aggregation, and recursion, as well as a host of declarative optimizations.  It also exploits the previously introduced semantic extensions for programs with \emph{aggregation in recursion}. 
As a result, the Spark programmer can now implement complex analytics pipelines of relational, graph and machine learning tasks in a single language, instead of stitching together programs written in different APIs, i.e., Spark SQL \cite{sparksql}, GraphX \cite{graphx} and MLlib. 


\subsection{Apache Spark}
Apache Spark \cite{zaharia2012resilient} is attracting a great deal of interest as a general platform for large-scale analytics, particularly because of its support for in-memory iterative analytics.  Spark enhances the MapReduce programming model by providing a language-integrated Scala API enabling the expression of programs as \emph{dataflows of second order transformations} (e.g., \texttt{map}, \texttt{filter}) on \emph{Resilient Distributed Datasets} (RDD) \cite{zaharia2012resilient}. An RDD is a distributed shared memory abstraction representing a partitioned dataset. RDDs are immutable, and transformations are coarse-grained and thus apply to all items in the RDD to produce a new RDD.  RDDs can be explicitly cached by the programmer in memory or on disk at workers.  RDDs provide fault tolerance by recomputing the sequence of transformations for the missing partition(s).

Once a Spark job is submitted, the scheduler groups transformations that can be pipelined into a single \emph{stage}. Stages are executed \emph{synchronously} in a topological order: a stage will not be scheduled until all stages it is dependent upon have finished  successfully.  Similar to MapReduce,  Spark \textit{shuffles} between stages to \textit{repartition} outputs among the nodes of the cluster. 
Spark has libraries for structured data processing (Spark SQL), stream processing (Spark Streaming), machine learning (MLlib), and graph processing (GraphX).  

\noindent \textbf{Spark as a runtime for Datalog.} Spark is a good candidate to support a Datalog compiler and Datalog evaluation; Spark is a general data processing system and provides the Spark SQL API \cite{sparksql}. Spark SQL provides logical and physical relational operators and Spark SQL's Catalyst compiler and optimizer supports the compilation and optimization of Spark SQL programs into physical plans.  \bigdatalogspark uses and extends Spark SQL operators, and also introduces operators implemented in the Catalyst framework so Catalyst planning features can be used on BigDatalog recursive plans.  

\bigdatalogspark is designed for general analytical workloads, and although we will focus much of the discussion and experiments on graph queries and recursive program evaluation, we do not claim that Spark is the best platform for graph workloads in general.  In fact, \bigdatalog\ can also be built into other general dataflow systems, including Naiad \cite{naiad2} and Hyracks \cite{hyracks}, and many of the optimization techniques presented in this section will also apply. 

\stitle{Challenges for Datalog on Spark.} The following represent the main challenges with implementing Datalog on Spark:

\begin{description}
\item[1. Spark SQL Supports Acyclic Plans:] 
Spark SQL lacks recursion operators, operators are designed for acyclic use, and the Catalyst optimizer plans non-recursive queries.

\item[2. Synchronous Scheduling:] Spark's synchronous stage-based scheduler requires unnecessary coordination for monotonic Datalog programs because monotonic Datalog programs are eventually consistent \cite{calmAmeloot,InterlandiT15}.

\item[3. Memory Utilization:] Each iteration of recursion will produce a new RDD to represent the updated recursive relation.  
If poorly managed, recursive applications on Spark can experience memory utilization problems.
\end{description}


\comment{
\paragraph{Benchmark Programs}
\label{sec:benchmark-programs}
In this paper, we focus on monotonic (positive) programs which include classical recursive queries from the literature as well as aggregate queries, some of which are long studied (e.g. shortest paths) and others studied more recently (connected components) \cite{seo2013distributed, myriadatalog}.

\vspace{0.5em}
\noindent \textbf{Classical Recursive Queries}
\vspace{-0.5em}
\begin{itemize}
  \setlength{\itemsep}{2pt}
  \setlength{\parskip}{0pt}
  \setlength{\parsep}{0pt}
  \item \textbf{Transitive Closure} (\tc)
  \item \textbf{Same Generation} (\sg) identifies pairs of vertices where both are the same number of hops from a common ancestor.
  \item \textbf{Reachability} (\bfs) produces all nodes connected by some path to a  given source node. 
\end{itemize}
\inv
\noindent \textbf{Aggregation in Recursion Queries}
\vspace{-0.5em}
\begin{itemize}
  \setlength{\itemsep}{2pt}
  \setlength{\parskip}{0pt}
  \setlength{\parsep}{0pt}
  \item \textbf{Single-Source Shortest Paths} (\sssp) computes the length of the shortest path from a source vertex to each vertex it is connected to.
  \item \textbf{Connected Components}  (\cc) identifies connected components in the graph.
\end{itemize}
}

\subsection{BigDatalog-Spark}\label{sec:bdexample}

We highlight the features of \bigdatalogspark with the help of the well known transitive closure (Example~\ref{ex:tc}) and same generation (Example~\ref{ex:sg}) programs.

\begin{example}[Transitive Closure (\tc)] \label{ex:tc}
\inv\bldl
\prule{r_1:tc(X,Y)}{arc(X,Y).}
\prule{r_2:tc(X,Y)}{tc(X,Z),\ arc(Z,Y).}
\edealrtl
\end{example}

\rone\ is an \textit{exit rule} because it serves as a base case of the recursion.  In \rone, the ${\tt arc}$ predicate represents the edges of the graph -- ${\tt arc}$ is a \textit{base relation}. \rone\ produces a ${\tt tc}$ fact for each ${\tt arc}$ fact. \rtwo\ will recursively produce ${\tt tc}$ facts from the conjunction of previously produced ${\tt tc}$ facts and ${\tt arc}$ facts.  The query to evaluate \tc is of the form ${\tt tc(X,Y)}$.  Lastly, this program uses a \textit{linear} recursion in \rtwo, since there is a single recursive predicate literal, whereas a \textit{non-linear} recursion would have multiple recursive literals in its body.  The number of iterations required to evaluate \tc is, in the worst case, equal to the longest simple path in the graph.

\begin{example}[Same Generation (\sg)] \label{ex:sg}
\inv\bldl
\prule{r_1:sg(X,Y)}{arc(P,X),\ arc(P,Y),\ X\ !=\ Y.}
\prule{r_2:sg(X,Y)}{arc(A,X),\ sg(A,B),\ arc(B,Y).}
\eldl
\end{example}

The exit rule \rone\ produces all ${\tt X,Y}$ pairs with the same parents (i.e. siblings) and the recursive rule \rtwo\ produces new ${\tt X,Y}$ pairs where both ${\tt X}$ and ${\tt Y}$ have parents of the same generation.

\bigdatalogspark programs are expressed as Datalog rules, then compiled, optimized and executed on Spark.  \figurename\ \ref{fig:apiexample} is the program to compute the size of the transitive closure of a graph using the \bigdatalogspark\ API.  The user first gets a \texttt{BigDatalogContext} (line 1), which wraps the \texttt{SparkContext} (\texttt{sc}) -- the entry point for writing and executing Spark programs. The user then specifies a schema definition for base relations and program rules (lines 2-4).  Lines 3-4 implement \tc\ from Example \ref{ex:tc}. The database definition and rules are given to the  \bigdatalogspark\ compiler which loads the database schema into a relation catalog (line 5). Next, the data source (e.g., local or HDFS file path, or RDD) for the \texttt{arc} relation is provided (line 6).  Then, the query to evaluate is given to the \texttt{BigDatalogContext} (line 7) which compiles it and returns an execution plan used to evaluate the query.  As with other Spark programs, evaluation is lazy -- the query is evaluated when \texttt{count} is executed (line 8).

\begin{figure}[!h]
\hrule
\vspace{0.125cm}
{\small
\begin{verbatim}
1  val bdCtx = new BigDatalogContext(sc)
2  val program = "database({arc(X:Integer, Y:Integer})."
3    + "tc(X,Y) <- arc(X,Y)."
4    + "tc(X,Y) <- tc(X,Z), arc(Z,Y)."
5  bdCtx.datalog(program)
6  bdCtx.datasource("arc", filePath)
7  val tc = bdCtx.query("tc(X,Y).")
8  val tcSize = tc.count()
\end{verbatim}}
\hrule
\caption{BigDatalog-Spark Program.}\label{fig:apiexample}
\vspace{-7ex}
\end{figure}

\paragraph{Parallel Semi-na{\"i}ve Evaluation on Spark.}\label{sec:psn}
\bigdatalogspark\ programs are evaluated using a parallel version of semi-na{\"i}ve evaluation we call {\it \psn\ evaluation} (PSN).  PSN is an execution framework for a recursive predicate and it is implemented using RDD transformations.  Since Spark evaluates synchronously, PSN will evaluate one iteration at a time; an iteration will not begin until all tasks from the previous iteration have completed.

The two types of rules for a recursive predicate -- the \textit{exit rules} and \textit{recursive rules} -- are compiled into separate \textit{physical plans} (plans) which are then used in the PSN evaluator.  Physical plans are composed of Spark SQL and \bigdatalogspark\ operators that produce RDDs.  
The exit rules plan is first evaluated once, and then the recursive rules plan is repeatedly evaluated until a fixpoint is reached.  Note that like the semi-na{\"i}ve evaluation, PSN will also evaluate symbolically rewritten rules (e.g., $\tt tc(X,Y) \leftarrow \delta tc(X,Z),\ arc(Z,Y).$).

\begin{algorithm}[!h]
\small
\begin{algorithmic}[1]
\caption{PSN Evaluator with RDDs}\label{alg:psnrdd}
\State $\mt delta = {\bf exitRulesPlan}.toRDD().distinct()$
\State $\mt all = delta$
\State $\mt updateCatalog(all, delta)$
\DoWhile
\State $\mt delta = {\bf recursiveRulesPlan}.toRDD()$
\State $\mt .subtract(all).distinct()$
\State $\mt all = all.union(delta)$
\State $\mt updateCatalog(all, delta)$
\EndDoWhile{$\mt (delta.{\bf count()} > 0)$}
\State $\mt {\bf return}\ all$
\end{algorithmic}
\end{algorithm}

Algorithm \ref{alg:psnrdd} is the pseudo-code for the PSN evaluator.  The  \textbf{exitRulesPlan} (line 1) and \textbf{recursiveRulesPlan} (line 5) are plans for the exit rules and recursive rules, respectively.  We use $\mt toRDD()$ (lines 1,5) to produce the RDD for the plan.  Each iteration produces two new RDDs -- an RDD for the new results produced during the iteration ($\mt delta$) and an RDD for all results produced thus far for the predicate ($\mt all$).  The $\mt updateCatalog$ (lines 3,8) stores new $\mt all$ and $\mt delta$ RDDs into a catalog for plans to access.  The exit rule plan is evaluated first.  The result is de-duplicated (\texttt{distinct}) (line 1) to produce the initial $\mt delta$ and $\mt all$ RDDs (line 2), which are used to evaluate the first iteration of the recursion.  Each iteration is a new job executed by  \texttt{count} (line 9).  First, the  \textbf{recursiveRulesPlan} is evaluated using the $\mt delta$ RDD from the previous iteration.  This will produce an RDD that is set-differenced (\texttt{subtract}) with the $\mt all$ RDD (line 6) and de-duplicated to produce a new $\mt delta$ RDD.  With lazy evaluation, the union of $\mt all$ and $\mt delta$ (line 7) from the previous iteration is evaluated prior to its use in \texttt{subtract} (line 6).

We have implemented PSN to cache RDDs that will be reused, namely $\mt all$ and $\mt delta$, but we omit this from Algorithm \ref{alg:psnrdd} to simplify its presentation.  Lastly, in cases of mutual recursion, when two or more rules belonging to different predicates reference each other (e.g., A $\leftarrow$ B, B $\leftarrow$ A), one predicate\footnote{Any of the mutually recursive predicates can be selected.} will ``drive'' the recursion with PSN and the other recursive predicate(s) will be an operator in the driver's recursive rules plan. 
%
The ``driver'' predicate is determined from the \emph{Predicate Connection Graph (PCG)}, which is basically a dependency graph constructed by the compiler. The use of PCG is common in many Datalog system architectures like $\cal LDL{++}$ \cite{ldl++}.

\subsection{Optimizations}\label{sec:opt-phys}
This section presents optimizations to improve the performance of \bigdatalogspark\ programs.  Details on the performance gain enabled by the discussed optimizations can be found in Tables 1 to 5 of the original \bigdatalogspark\ paper~\cite{bigdatalog}.

\paragraph{Optimizing PSN.}\label{sec:setrdd}
As shown with Algorithm \ref{alg:psnrdd}, PSN can be implemented with RDDs and standard transformations. However, using standard RDD transformations is inefficient because at each iteration the results of the recursive rules are set-differenced with the entire recursive relation (line 6 in Algorithm \ref{alg:psnrdd}), which is growing in each iteration, and thus expensive data structures must be created for each iteration.  We propose, instead, the use of \emph{SetRDD}, which is a specialized RDD for storing distinct \texttt{Row}s and tailored for set operations needed for PSN.  Each partition of a SetRDD is a set data structure.
Although an RDD is intended to be immutable, we make SetRDD mutable under the union operation.  The \texttt{union} mutates the set data structure of each SetRDD partition and outputs a new SetRDD comprised of these same set data structures.  If a task performing union fails and must be re-executed, this approach  will not lead to incorrect results because union is monotonic and facts can be added only once.  
Lastly, SetRDD transformations are implemented to not shuffle, and therefore the compiler must add shuffle operators to a plan.  This approach allows for a simplified and generalized PSN evaluator. 


\comment{
\begin{table}[!b]
\centering
\caption{PSN vs. PSN with SetRDD Performance} \label{tab:psnsetrddresults}
\scriptsize
{\renewcommand{\arraystretch}{1.1}%
{\setlength{\tabcolsep}{0.4em}
\begin{tabular}{|l|r|r|r|r|r|r|}
 \hline
 \multirow{2}{*}{Time (s)} & \multicolumn{3}{c|}{\tc} & \multicolumn{3}{c|}{\sg} \\
 \cline{2-7}
 & \treeseventeen & \gridone & \gtenk & \treeeleven & \gridone & \gtenk \\
 \hline
 PSN & 244 & OOM & 208 & OOM & 230 & 1129\\
 PSN with SetRDD & 41 & 134 & 20 & 59 & 61 & 130\\
 \hline
\end{tabular}}}
\end{table}}

\begin{figure}[!h]
\centering
 \begin{subfigure}[b]{0.30\textwidth}
    \centering
    \includegraphics[width=\textwidth]{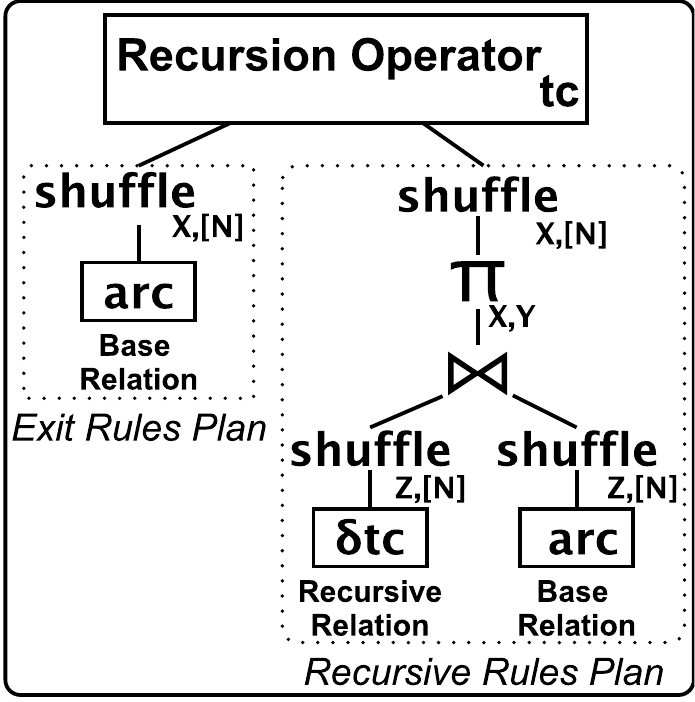}
    \caption{\tc} \label{fig:physicalplanstc1st}
  \end{subfigure}%
  \hspace{1em}
  \begin{subfigure}[b]{0.38\textwidth}
    \centering
    \includegraphics[width=\textwidth]{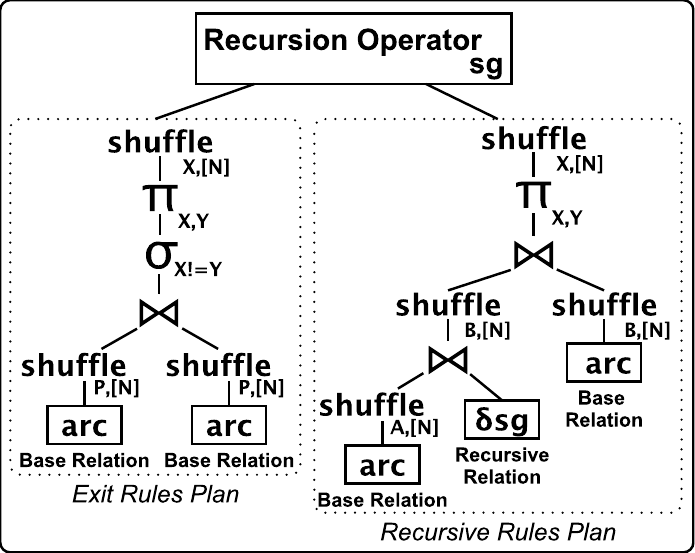}
    \caption{\sg} \label{fig:physicalplanssg1st}
  \end{subfigure}%
  \caption{PSN with SetRDD Physical Plans.}\label{fig:psnsetrddplans}
\end{figure}

\paragraph{Partitioning.}
An earlier research on Datalog\ showed that a good partitioning strategy 
(i.e., finding the arguments on which to partition) for a recursive predicate 
was important for an efficient parallel 
 evaluation \cite{Cohen:1989:WSP:73721.73742,DBLP:conf/sigmod/GangulyST90,DBLP:journals/jlp/GangulyST92,DBLP:conf/sigmod/WolfsonO90}. Since transferring data (i.e., communication) has a high cost in a cluster, we seek a partitioning strategy that limits shuffling.  The default partitioning strategy employed by \bigdatalogspark\ is to partition the recursive predicate on the \textit{first argument}.  
\figurename\ \ref{fig:psnsetrddplans}(\subref{fig:physicalplanstc1st}) is the plan for \examplename\ \ref{ex:tc} for PSN with SetRDD.  With the recursive predicate ($\tt tc$) partitioned on the first argument both the exit rule and recursive rule plans terminate with a shuffle operator.

In the plan in \figurename\ \ref{fig:psnsetrddplans}(\subref{fig:physicalplanstc1st}) $\tt \delta tc$ requires  shuffling prior to the join since it is not partitioned on the join key ($\tt Z$) because the default partitioning is the first argument ($\tt X$).  However, if the default partitioning strategy was to use instead the second argument, the inefficiency with \figurename\ \ref{fig:psnsetrddplans}(\subref{fig:physicalplanstc1st}) would be resolved but then other programs such as \sg\ (plan shown in \figurename\ \ref{fig:psnsetrddplans}(\subref{fig:physicalplanssg1st})) would suffer ($\tt \delta sg$ would require a shuffle prior to the join).  Therefore, \bigdatalogspark\ allows the user to define a recursive predicate's partitioning via configuration.  

\comment{
\begin{table}[h!]
\centering
\caption{Comparison of \tc\ with Different Partitioning} \label{tab:leftvsrighttcresults}
\scriptsize
{\renewcommand{\arraystretch}{1.1}%
\begin{tabular}{|l|r|r|r|r|}
 \hline
Time (s) & \treeseventeen & \gridtwo & \gtenk \\
 \hline
1st Argument & 41 & 370 & 20\\
2nd Argument & 26 & 265 & 19\\
 \hline
\end{tabular}}
\end{table}

\begin{table}[h]
\centering
\caption{Join Optimizations for Linear Recursion} \label{tab:joinresults}
\scriptsize
{\renewcommand{\arraystretch}{1.1}%
{\setlength{\tabcolsep}{0.4em}
\begin{tabular}{|l|r|r|r|r|}
 \hline
 \multirow{2}{*}{Time (s)} & \multicolumn{2}{c|}{\tc} & \multicolumn{2}{c|}{\sg} \\
 \cline{2-5}
 & \treeseventeen & \gridtwo & \treeeleven & \gridtwo \\
 \hline
 Shuffle join no caching & 26 & 265 & 59 & 107 \\
 Shuffle join caching & 17 & 196 & 56 & 81 \\
 Broadcast join& 53 & 197 & 45 & 54 \\
 \hline
\end{tabular}}}
\end{table}
}

\paragraph{Join Optimizations for Linear Recursion.}
By keeping the number of partitions static, a \textit{shuffle join} implementing a linear recursion can have the non-recursive join input cached because the non-recursive inputs will not change during evaluation.  This can lead to significant performance improvement since input partitions no longer have to be shuffled and loaded into lookup tables prior to the join in each iteration.  

\begin{figure}[!h]
\captionsetup{justification=centering}
  \begin{minipage}{.40\textwidth}
    \centering
    \includegraphics[scale=0.52]{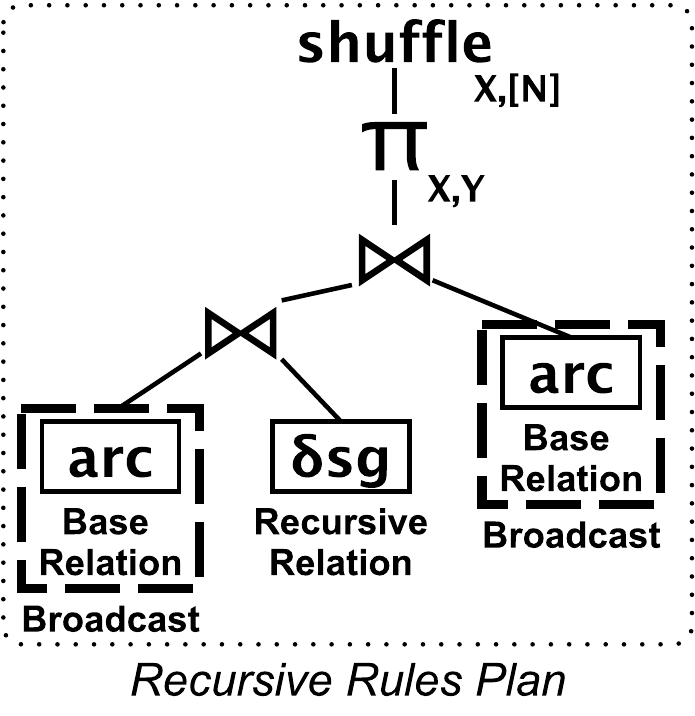}
    \caption{\sg\ with Broadcast Joins.} \label{fig:sgbroadcast}
  \end{minipage}%
\hspace{10ex}
  \begin{minipage}{.40\textwidth}
    \centering
    \includegraphics[scale=0.535]{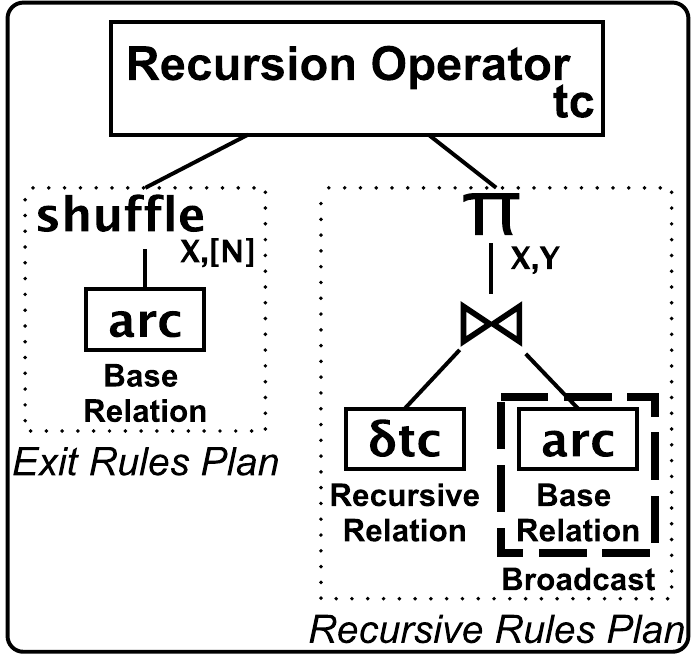}
    \vspace{0.5em}
    \captionof{figure}{Decomposable \tc\ Plan.}
  \label{fig:decomposabletc}
\end{minipage}
\end{figure}

Instead of shuffle joins, each partition of a recursive relation can be joined with an entire relation (e.g., \textit{broadcast join}).  For either type of join, the non-recursive input is loaded into a lookup table.  For a broadcast join, the cost of loading the entire relation into a lookup table is amortized over the recursion because the lookup table is cached and then reused in every iteration.  \figurename\ \ref{fig:sgbroadcast} shows a recursive rules plan for Example \ref{ex:sg} (\sg) with two levels of broadcast joins.  In the event that a broadcast relation is used multiple times in a plan, as in \figurename\ \ref{fig:sgbroadcast}, \bigdatalogspark\ will broadcast it once and share it among all broadcast join operators joining the relation.


\paragraph{Decomposable Programs.}\label{sec:decomposable}
Previous research on parallel evaluation of Datalog programs determined that some programs are \textit{decomposable} and thus evaluable in parallel without redundancy (a fact is only produced once) and without processor communication or synchronization \cite{DBLP:conf/sigmod/WolfsonS88}.  
Since mitigating the cost of synchronization and shuffling can lead to significant execution time speedup, enabling \bigdatalogspark\ to support techniques for identifying and evaluating decomposable programs is desirable. 

We consider a \bigdatalogspark\ physical plan decomposable if the recursive rules plan has no shuffle operators. Example \ref{ex:tc} (linear \tc) is a decomposable program \cite{DBLP:conf/sigmod/WolfsonS88} however, its physical plan shown in \figurename\ \ref{fig:psnsetrddplans}(\subref{fig:physicalplanstc1st}) has shuffle operators in the recursive rules plan.  Instead, \bigdatalogspark\ can produce a decomposable physical plan for Example \ref{ex:tc}.  First, $\tt tc$ will be partitioned by its first argument which divides the recursive relation so that each partition can be evaluated independently and without shuffling. Secondly, a broadcast join will be used which allows each partition of the recursive relation to join with the \textit{entire} $\tt arc$ base relation.  \figurename\ \ref{fig:decomposabletc} is the decomposable physical plan for Example \ref{ex:tc}.  Base relations are not pre-partitioned, therefore the exit rules plan has a shuffle operator to repartition the ${\tt arc}$ base relation by ${\tt arc}$'s first argument $\tt X$ into $\tt N$ partitions.  

\comment{
\begin{table}[h]
\centering
\caption{Shuffle vs. Decomposable TC Plans} \label{tab:decomposableresults}
\scriptsize
{\renewcommand{\arraystretch}{1.1}%
{\setlength{\tabcolsep}{0.4em}
\begin{tabular}{|l|r|r|r|r|r|}
 \hline
 Time (s) & \treeseventeen & \gridtwo & \gtenk & \gtenkzeroone & \gtwentyk \\
 \hline
 Shuffle &  26 & 265  & 19 & 121 & 101\\
 Decomposable & 49  & 55 & 7  & 22 & 19 \\
 \hline
\end{tabular}}}
\end{table}
}

\bigdatalogspark\ identifies decomposable programs via syntactic analysis of program rules using techniques presented in the \textit{generalized pivoting} work \cite{generalizedpivoting}.  The authors of \cite{generalizedpivoting} show that the existence of a \textit{generalized pivot set} (GPS) for a program is a sufficient condition for decomposability and present techniques to identify GPS in arbitrary Datalog programs.  
When a \bigdatalogspark\ program is submitted to the compiler, the compiler will apply the generalized pivoting solver to determine if the program's recursive predicates have GPS.  If they indeed have, we now have a partitioning strategy and in conjunction with broadcast joins we can efficiently evaluate the program with these settings.  For example, Example \ref{ex:tc} has a GPS which says to partition the $\tt tc$ predicate on its first argument.  Note that this technique is enabled by using Datalog and it allows \bigdatalogspark\ to analyze the program at the logical level.  The Spark API alone is unable to provide this support since programs are written in terms of physical operations.

\comment{
\subsection{Aggregates}\label{sec:agg}
\bigdatalogspark\ supports non-monotonic aggregates (e.g., traditional SQL aggregates) \texttt{min}, \texttt{max}, \texttt{sum}, \texttt{count}, \texttt{avg}. As an example, consider Example \ref{ex:trianglecounting}, the \bigdatalogspark\ program which counts the triangles in a graph, an important program in network analysis.  In this non-recursive program, \rone\ performs self-joins of ${\tt arc}$ to produce triangle occurrences which are then counted by \rtwo.  
Lastly, note that although this program is expressed as two Datalog\ rules, this program is a 50+ line GraphX program. 

\comment{
\begin{example}{Triangle Counting} \label{ex:trianglecounting}
\inv\bdealrtl
\prulertlt{r1}{triangles(X,Y,Z)}{arc(X,Y),X<Y,arc(Y,Z),Y<Z,arc(Z,X).}
\prulertlt{r2}{count\_triangles(\countv{\_})}{triangles(X,Y,Z).}
\edealrtl
\end{example}
}

However, non-monotonic aggregates cannot be used in recursion. As described in Section~\ref{sec:extrema}, we have recently proposed a rewrite technique and aggregates that are monotonic w.r.t. set containment, the same monotonicity used by standard Datalog, meaning these aggregates can be used in recursive rules and evaluated using techniques such as SN and magic sets \cite{mazuran2013declarative,mazuran2012extending}.  
We have presented a sequential version of these aggregates in \cite{shkapsky2015optimizing}, whereas \bigdatalogspark implements a distributed version of the aggregates.  

\bigdatalogspark\ supports four monotonic aggregates - \mmin, \mmax, \mcnt, \msum.  The declarative semantics allows the aggregates inside the recursion so long as monotonicity w.r.t. set containment is maintained.  Therefore, during evaluation the monotonic aggregates can produce new higher (\mmax, \mcnt, \msum) or lower (\mmin) values with each input fact and thus an outer non-monotonic aggregate ($\tt min$ or $\tt max$) is necessary to produce only the final value. Additionally, the rewrites described in Section~\ref{sec:extrema} can be employed. 

\comment{\begin{example}{Single-Source Shortest Paths} \label{ex:sssp}
\inv\bdealrtl
\prulertlt{r1}{sssp2(Y,\mminv{D})}{Y=1, D=0.}
\prulertlt{r2}{sssp2(Y,\mminv{D})}{sssp2(X,D1), arc(X,Y,D2), D=D1+D2.}
\prulertlt{r3}{sssp(X,\minv{D})}{sssp2(X,D).}
\edealrtl
\end{example}

The \sssp\ program computes the length of the shortest path from a source vertex to all vertices it is connected to.  This program uses a \mmin\ monotonic aggregate.  Here the \texttt{arc} predicate in \rtwo\ denotes edges of the graph ($\tt X,Y$) with edge cost $\tt D2$.  \rone\ seeds the recursion with starting vertex 1.  Then, \rtwo\ will recursively produce all new minimum cost paths to a node \texttt{Y} though node \texttt{X}.  Lastly, \rthree\ produces only the minimum cost path for each node \texttt{X}, however in our actual implementation, we do not have to evaluate \rthree\ since at the completion of the recursion, $\tt sssp2$'s relation will contain the shortest path from 1 to each vertex.
}

\stitle{Evaluation and Implementation.} Programs with monotonic aggregates in recursive rules are evaluated with an aggregate version of PSN we call \textit{Parallel Semi-naive - Aggregate} (PSN-A).  Compared with PSN, PSN-A is a simpler evaluator.  Since new facts are only produced when a greater (\mmax, \mcnt, \msum) or lesser (\mmin) value than the previous value for the (aggregate) group is produced, de-duplication is unnecessary.  Furthermore, the union is unnecessary because new results are added to the aggregate relation during aggregate evaluation.  We implement PSN-A in an aggregate recursion operator.  Also, we use a specialized RDD called an \emph{AggregateSetRDD}, in which each partition is a key value map where each entry represents a unique group and its current value.  Caching AggregateSetRDD avoids the expense of reloading key value maps each iteration for aggregate.  Additionally, since the aggregate functions are monotonic, as with SetRDD's \texttt{union} operation, AggregateSetRDD is mutable under aggregate evaluation.  AggregateSetRDD will reference the same maps as its creator.  Should a task fail during evaluation, any changes to the aggregate partition will not result in incorrect results since a value can only be updated if it is higher (\mmax, \mcnt, \msum) or lower (\mmin) than previously computed values.
}

\subsection{Experiments}\label{sec:perf}

\begin{figure*}[t]
\centering
\includegraphics[width=\textwidth]{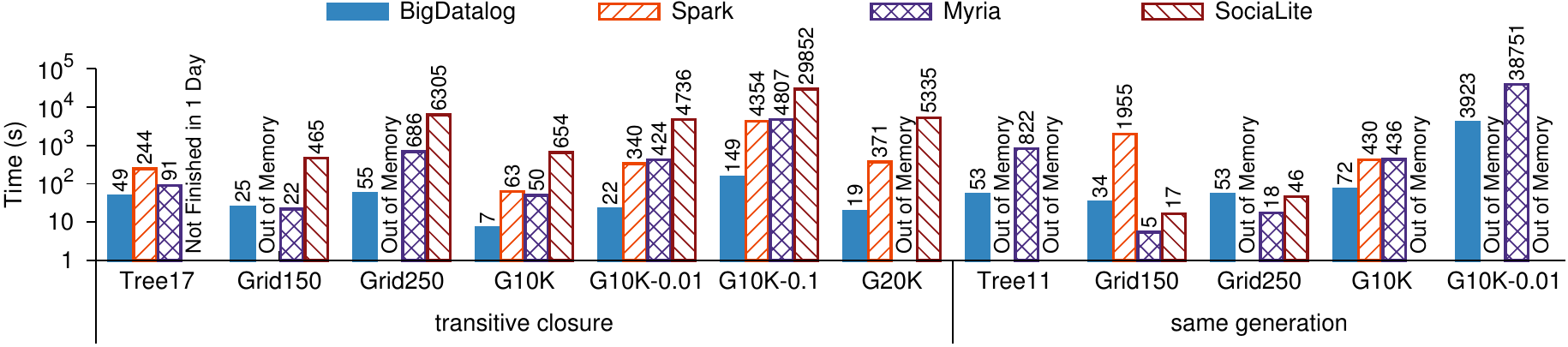}
\caption{System Comparison using \tc and \sg.}
\label{fig:systemcomparions}
\vspace{-1ex}
\end{figure*}

\begin{figure}[h!]
\centering
\begin{subfigure}[t]{0.4\textwidth}
\centering
\includegraphics[width=\textwidth]{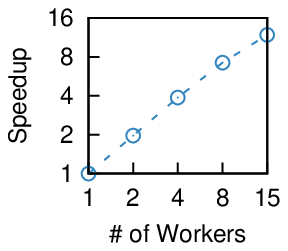}
\vspace{-1.5em}
\caption{\tc on \texttt{\gtwentyk}}\label{fig:tcscaleout}
\end{subfigure}
\begin{subfigure}[t]{0.4\textwidth}
\centering
\includegraphics[width=\textwidth]{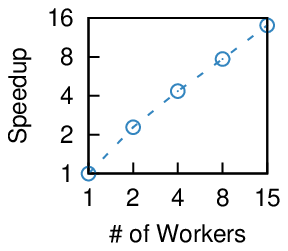}
\vspace{-1.5em}
\caption{\sg on \texttt{\gtenk}}\label{fig:sgscaleout}
\end{subfigure}
\caption{Scaling-out Cluster Size.}\label{fig:scaleout}
\vspace{-2ex}
\end{figure}

\begin{figure}[h!]
\centering
\begin{subfigure}[t]{0.44\textwidth}
\centering
\includegraphics[width=\textwidth]{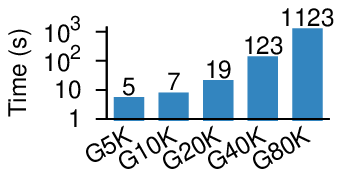}
\vspace{-1.5em}
\caption{\tc}\label{fig:tcscaleup}
\end{subfigure}
\begin{subfigure}[t]{0.4\textwidth}
\centering
\includegraphics[width=\textwidth]{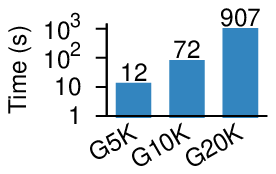}
\vspace{-1.5em}
\caption{\sg}\label{fig:sgscaleup}
\end{subfigure}
\caption{Scaling-up on Random Graphs.}\label{fig:scaleup}
\end{figure}

In~\cite{bigdatalog} we have tested \bigdatalogspark over both synthetic and real-world datasets, and compared against other distributed Datalog implementations (e.g., Myria~\cite{myria} and SocialLite~\cite{seo2013distributed}), as well as hand-coded versions of programs implemented in Spark. The tests were executed using the \tc, \sg, \cc, \pymk, and \mlm programs presented in Section~\ref{sec:graph} plus some additional ones.  Here showcase a systems comparison using \tc and \sg (Figure~\ref{fig:systemcomparions}) and discuss results of scale-out and scale-up experiments (respectively in Figure~\ref{fig:scaleout} and Figure~\ref{fig:scaleup}). Each execution time reported in the figures is calculated by performing the same experiment five times, discarding the highest and lowest values, and taking the average of the remaining three values. The unit of time measurement is seconds.

\stitle{Configuration.} Our experiments were run on a 16 node cluster. Each node ran Ubuntu 14.04 LTS and had an Intel i7-4770 CPU (3.40GHz, 4 core/8 thread), 32GB memory and a 1 TB 7200 RPM hard drive. Nodes were connected with 1Gbit network. The \bigdatalogspark implementation ran on Spark 1.4.0 and the file system is Hadoop 1.0.4.

\stitle{Datasets.} Table \ref{tab:benchmarkdatasynth} shows the synthetic graphs used for the experiments of this section and of Section~\ref{sec:datalog_multicore}.
  \texttt{Tree11} and \texttt{Tree17} are trees of height 11 and 17 respectively, and the degree of a non-leaf vertex is a random number between 2 and 6.  \texttt{Grid150} is a 151 by 151 grid while \texttt{Grid250} is a 251 by 251 grid.  The G\textit{n}-{\textit{p}} graphs are \textit{n}-vertex random graphs (Erd\H{o}s-R\'enyi model) generated by randomly connecting vertices so that each pair is connected with probability \textit{p}.  G\textit{n}-\textit{p} graph names omitting \textit{p} use default probability 0.001.  Note that for these graphs \tc\ and \sg\ are capable of producing result sets many orders of magnitude larger than the input dataset, as shown by the last two columns in Table~\ref{tab:benchmarkdatasynth}.

\begin{table*}[h]
\centering
\caption{Parameters of Synthetic Graphs}\label{tab:benchmarkdatasynth}
\begin{minipage}[t]{0.85\linewidth}
\begin{tabular}{|l|r|r|r|r|}
\hline
\textbf{Name} & \textbf{Vertices} & \textbf{Edges} & \textbf{\tc} & \textbf{\sg}\\
\treeeleven & 71,391 & 71,390 & 805,001 & 2,086,271,974\\
\treeseventeen & 13,766,856 & 13,766,855 & 237,977,708 & \rule[0.5ex]{6em}{0.4pt}\\
\gridone & 22,801 & 45,300 & 131,675,775 & 2,295,050\\
\gridtwo & 63,001 & 125,500 & 1,000,140,875 & 10,541,750\\
\gfivek & 5,000 & 24,973 & 24,606,562 & 24,611,547\\
\gtenk & 10,000 & 100,185 & 100,000,000 & 100,000,000\\
\gtenkzeroone & 10,000 & 999,720 & 100,000,000 & 100,000,000\\
\gtenkone & 10,000 & 9,999,550 & 100,000,000 & 100,000,000\\
\gtwentyk & 20,000 & 399,810 & 400,000,000 & 400,000,000\\
\gfourtyk & 40,000 & 1,598,714 & 1,600,000,000 & 1,600,000,000\\
\geightyk & 80,000 & 6,399,376 & 6,400,000,000 & 6,400,000,000\\
\hline
\end{tabular}
\end{minipage}
\end{table*}

\stitle{Systems Comparison.}
For \tc, \bigdatalogspark\ uses \examplename\ \ref{ex:tc} with the decomposed plan from \figurename\ \ref{fig:decomposabletc}. 
For \sg, \bigdatalog uses \examplename\ \ref{ex:sg} with broadcast joins (\figurename\ \ref{fig:sgbroadcast}).  We use equivalent programs in Myria and Socialite, and hand-optimized semi-na{\"i}ve programs written in the Spark API which are implemented to minimize shuffling. 
\figurename\ \ref{fig:systemcomparions} shows the evaluation time for all four systems. 

\bigdatalogspark\ is the only system that finishes the evaluation for \tc\ and \sg\ on all graphs except \sg\ on \treeseventeen\ since the size of the result is larger than the total disk space of the cluster. \bigdatalogspark\ has the fastest execution time on six of the seven graphs for \tc; on four of the graphs it outperforms the other systems by an order of magnitude.
The \bigdatalogspark\ plan only performs an initial shuffle of the dataset, and then evaluates the recursion without shuffling, and proves very efficient.
In the case of \gridone, which is the smallest graphs used in this experiment in terms of both edges and queries output sizes, Myria outperforms \bigdatalogspark both in \tc and \sg. This is explained as the evaluation requires many iterations, where each iteration performs very little work, and therefore the overhead of scheduling in \bigdatalogspark\ takes a significant portion of execution time.
However, as the data set becomes larger the superior scalability of \bigdatalogspark comes into play enabling it to outperform all other systems on \gridtwo. In fact, \figurename\ \ref{fig:systemcomparions} shows that the execution time of \bigdatalogspark on \tc only grows to 2.2 times those of \gridone, whereas those of Myria and Socialite grow by more than one order of magnitude; from \gridone\ to \gridtwo, \bigdatalogspark also scales better on \sg compared to the other systems.
The Spark programs are not only affected by the overhead of scheduling and shuffling, but also suffer memory utilization issues related to dataset caching, and therefore ran out of memory for several datasets both in \tc and \sg.

\stitle{Scalability.}
In this set of experiments  we use the G\textit{n}-{\textit{p}} graphs.   \figurename\ \ref{fig:scaleout}(\subref{fig:tcscaleout}) shows the speedup for \tc\ on \gtwentyk\ as the number of workers increases from one to 15 (all with one master) \textit{w.r.t.} using only one worker, and \figurename\ \ref{fig:scaleout}(\subref{fig:sgscaleout}) shows the same experiment run for \sg\ with \gtenk.  Both figures show a linear speedup, with the speedup of using 15 workers as 12X and 14X for \tc\ and \sg, respectively.

The scaling-up results shown in \figurename\ \ref{fig:scaleup} were ran with the full cluster, i.e., one master and 15 workers. With each successively larger graph size, i.e., from \gfivek\ to \gtenk, the size of the transitive closure quadruples, but we do not observe a quadrupling of the evaluation time. Instead, evaluation time increases first less than 1.5X (\gfivek\ to \gtenk), then 3X (\gtenk\ to \gtwentyk), 6X (\gtwentyk\ to \gfourtyk), and 9X (\gfourtyk\ to \geightyk). Rather than focusing on the size of the \tc\ w.r.t. execution time, the reason for the increase in execution time is explained by examining the results in Table~\ref{tab:tcscalingdetails}.

\begin{table*}[h]
\centering
\caption{~\tc ~Scaling-up Experiments Result Details\label{tab:tcscalingdetails}}
\inv{\small (Execution time not including the time to broadcast $\tt arc$)}
\begin{minipage}[t]{\linewidth}
{\setlength{\tabcolsep}{0.4em}
\begin{tabular}{|l|r|r|r|r|r|}
\hline
\multirow{2}{*}{\textbf{Graph}} & \textbf{Time -} & \multirow{2}{*}{\textbf{\tc}} & \multirow{2}{*}{\textbf{Generated Facts}} & \textbf{Generated} & \textbf{Generated} \\
& \textbf{broadcast(s)} & & & \textbf{Facts / \tc} & \textbf{Facts / Sec.} \\
\gfivek & 4 & 24,606,562 & 122,849,424 & 4.99 & 30,712,356 \\
\gtenk  & 6 & 100,000,000 & 1,001,943,756 & 10.02 & 166,990,626 \\
\gtwentyk & 17 & 400,000,000 & 7,976,284,603 & 19.94 & 469,193,212 \\
\gfourtyk & 119 & 1,600,000,000 & 50,681,485,537 & 31.68 & 425,894,836 \\
\geightyk & 1112 & 6,400,000,000 & 510,697,190,536 & 79.80 & 459,673,439 \\
\hline
\end{tabular}
}
\end{minipage}
\end{table*}

Broadcasting the $\tt arc$ relation requires between one second for \gfivek\ to twelve seconds for \geightyk. Table~\ref{tab:tcscalingdetails} shows the execution time minus the time to broadcast $\tt arc$, which is the total time the program required to actually evaluate \tc. Table~\ref{tab:tcscalingdetails} also shows the number of generated facts, which is the number of total facts produced prior to de-duplication and is representative of the actual work the system must perform to produce the \tc (i.e., HashSet lookups), the ratio between \tc\ size and generated facts, and the number of generated facts per second (time - broadcast time), which should be viewed as the evaluation throughput.  These details help to explain why the execution times increase at a rate greater than the increase in \tc\ size -- the number of generated facts is increasing at a rate much greater than the increase in \tc\ size.  The last column shows that even with the increase in number of generated facts, \bigdatalogspark\ still maintains good throughput throughout.  Continuing, the first two graphs are too small to stress the system, but once the graph is large enough (e.g., \gtwentyk) the system exhibits a much greater throughput, which is stable across the larger graphs.

\begin{table*}[h]
\centering
\caption{\sg Scaling-up Experiments Result Details}\label{tab:sgscalingdetails}
\inv{\small (Execution time not including the time to broadcast $\tt arc$)}
\begin{minipage}[t]{\linewidth}
{\setlength{\tabcolsep}{0.4em}
\begin{tabular}{|l|r|r|r|r|r|}
\hline
\multirow{2}{*}{\textbf{Graph}} & \textbf{Time -} & \multirow{2}{*}{\textbf{\sg}} & \multirow{2}{*}{\textbf{Generated Facts}} & \textbf{Generated} & \textbf{Generated} \\
& \textbf{broadcast(s)} & & & \textbf{Facts / \sg} & \textbf{Facts / Sec.} \\
\gfivek & 11 & 24,611,547 &	612,891,161	& 24.90 & 55,717,378 \\
\gtenk & 71 & 100,000,000 &	10,037,915,957 &	 100.38 & 141,379,098 \\
\gtwentyk & 905 & 400,000,000 & 159,342,570,063 & 398.36 & 176,069,138 \\
\hline
\end{tabular}
}
\end{minipage}
\end{table*}

Table~\ref{tab:sgscalingdetails} displays the same details as Table~\ref{tab:tcscalingdetails} but for \sg.  Table~\ref{tab:sgscalingdetails} displays the execution time-minus the broadcast time of $\tt arc$, the result set size, the number of generated facts as well as statistics for the ratio of generated facts for each \sg\ fact and generated fact per second of evaluation (throughput).  With \sg, the number of generated facts is much higher than we observe with \tc, reflecting the greater amount of work \sg\ requires.  For example, on \gtenk\ and \gtwentyk\ \sg\ produces 10X and 20X the number of generated facts, respectively, than \tc\ produces.  We also observe a much greater rate of increase in generated facts between graph sizes for \sg\ compared to \tc.  For example, from \gtenk\ to \gtwentyk\ we see a 16X increase in generated facts for \sg\ versus only an 8X increase for \tc.  For \sg, we do not achieve as high a throughput as with \tc, which is explained in part by the fact that \sg\ requires shuffling, whereas our \tc\ program evaluates purely in main memory after an initial shuffle.
\section{Datalog on Multicore Systems: \bigdatalogmc}
\label{sec:datalog_multicore}

Multicore machines are composed by one or more processors, where each of them contains several cores on a chip~\cite{DBLP:journals/corr/abs-1110-3535}. Each core is composed of computation units and caches, while the main memory is commonly shared. While the individual cores do not necessarily run as fast as the highest performing single-core processors, they are able to obtain high performance by handling more tasks in parallel.
In this paper we will consider multicore processors implemented as a group of \emph{homogeneous cores}, where the same computation logic is applied in a divide-and-conquer way over a partition of the input dataset. 

Unfortunately, single-core applications do not get faster automatically on a multicore architecture with the increase of cores. For this reason, programmers are forced to write 
specific parallel logic to exploit the performance of multicore architectures. 
Next we present the techniques used by \bigdatalogmc to enable the efficient parallel evaluation of Datalog programs over a shared-memory multicore machine with $n$ processors.

\subsection{Parallel Bottom-Up Evaluation}

We start with how \bigdatalogmc performs the parallel bottom-up evaluation of the transitive closure program \tc in \cexp\ref{ex:tc}. We divide each relation into $n$ partitions and we use the relation name with a superscript $i$ to denote the $i$-th partition of the relation. Each partition has its own storage for tuples, unique index, and secondary indexes. Assuming that there are $n$ workers that perform the actual query evaluation, and one coordinator that manages the coordination between the workers. \cexp\ref{example:tc-parallel} below shows a parallel evaluation plan for \tc.

\begin{example}[Parallel bottom-up evaluation of \tc]\label{example:tc-parallel}
Let $h$ be a hash function that maps a vertex to an integer between 1 to $n$. Both ${\tt arc}$ and ${\tt tc}$ are partitioned by the first column, i.e., $h({\tt X}) = i$ for each ${\tt (X, Y)}$ in ${\tt arc}^i$ and $h({\tt X}) = i$ for each ${\tt (X, Y)}$ in ${\tt tc}^i$. 
The parallel evaluation proceeds as follows.
\begin{enumerate}
\item The $i$-th worker evaluates the exit rule by adding a tuple ${\tt (X, Y)}$ to ${\tt tc}$ for each tuple ${\tt (X, Y)}$ in ${\tt arc}^i$.
\item Once all workers finish Step (1), the coordinator notifies each worker to start Step (3).
\item For each new tuple ${\tt (X, Z)}$ in ${\tt tc}^i$ derived in the previous iteration, the $i$-th worker looks for tuples of the form ${\tt (Z, Y)}$ in ${\tt arc}$ and adds a tuple ${\tt (X, Y)}$ to ${\tt tc}$.
\item Once all workers finish Step (3), the coordinator checks if the evaluation for ${\tt tc}$ is completed. If so, the evaluation terminates; otherwise, the evaluation starts from Step (3).
\end{enumerate}
In Step (1) and Step (3), each worker performs its task on one processor while the coordinator waits. Step (2) and Step (4) serve as synchronization barriers.
\end{example}

In the above example, the $i$-th worker only writes to ${\tt tc}^i$ in Step (1), and it only reads from and writes to ${\tt tc}^i$ in Step (3). Thus, ${\tt tc}^i$ is only accessed by the $i$-th worker. This property does not always hold in every evaluation plan of ${\tt tc}$. For example, if we keep the current partitioning for ${\tt arc}$ but instead partition ${\tt tc}$ by its second column, then every worker could write to ${\tt tc}^i$ in Step (3), and multiple write operations to ${\tt tc}^i$ can occur concurrently; in this plan, we use a lock to ensure only one write operation to ${\tt tc}^i$ is allowed at a time---a worker needs to acquire the lock before it writes to ${\tt tc}^i$, and it releases the lock once the write operation completes.

In general, we use a lock to control the access to a partition if multiple read/write operations can occur concurrently. There are two types of locks: (i) an {\em exclusive lock} (x-lock) that allows only one operation at a time; and (ii) a {\em readers--writer lock} (rw-lock) that a) allows only one write operation at a time, b) allows concurrent read operations when no write operation is being performed, and c) disallows any read operation when a write operation is being performed. We use (i) an x-lock if there is no read operation and only multiple write operations can occur concurrently; (ii) an rw-lock if multiple read and write operations can occur concurrently since it allows for more parallelism than an x-lock.

We assume that
every relation is partitioned using the same hash function $h$ defined as
\begin{displaymath}
h(x_1, \ldots, x_t) = \sum_{i=1}^t g(x_i) \bmod n,
\end{displaymath}
where the input to $h$ is a tuple of any arity $t$ and $g$ is a hash function with a range no less than $n$. 
Then the key factor that 
determines whether locks are required during the evaluation
is how each relation is partitioned, 
which is specified
using {\em discriminating sets}. A discriminating set of a (non-nullary) relation $R$ is a non-empty subset of columns in $R$. Given a discriminating set of a relation, we divide the relation into $n$ partitions by the hash value of the columns that belong to the discriminating set. For each predicate $p$ that corresponds to a base relation or a derived relation, let $R$ be the relation that stores all tuples corresponding to facts about $p$ in memory; we select a discriminating set of $R$ that specifies the partitioning of $R$ used in the evaluation of $p$. The collection of all the selected discriminating sets uniquely determines how each relation is partitioned. These discriminating sets can be arbitrarily selected as long as there is a unique discriminating set for each derived relation.

\begin{example}[Discriminating sets for the plan in \cexp\ref{example:tc-parallel}]
The discriminating sets for the two occurrences of ${\tt arc}$ are both $\{1\}$. Moreover, ${\tt tc}$ is a derived relation, and its discriminating set is $\{1\}$.
\end{example}

\subsection{Parallel Evaluation of AND/OR Trees}\label{sec:andortree}
The internal representation used by \bigdatalogmc to represent a Datalog program is an AND/OR tree \cite{ldl++}. An OR node represents a predicate and an AND node represents the head of a rule. The root is an OR node. The children of an OR node (resp., AND node) are AND nodes (resp., OR nodes). Each node has a ${\tt getTuple}$ method that calls the ${\tt getTuple}$ methods of its children. Each successful invocation to the method instantiates the variables of one child (resp., all the children) and the parent itself for an OR node (resp., AND node). The program is evaluated by repeatedly applying the ${\tt getTuple}$ method upon its root until it fails. Thus, for an OR node, the execution (i) exhausts the tuples from the first child; (ii) continues to the next child; and (iii) fails when the last child fails. An OR node is an {\em R-node} if it reads from a base or derived relation with its ${\tt getTuple}$ method, while 
it is a {\em W-node} if it writes to a derived relation with its ${\tt getTuple}$ method. 
Finally, an OR node is an {\em entry node} if (i) it is a leaf, and (ii) it is the first R-node among its siblings, and (iii) none of its ancestor OR nodes has a left sibling (i.e., a sibling that appears before the current node) that has an R-node descendant or a W-node descendant.

\begin{figure}[htbp]
\centering
\includegraphics[width=0.8\textwidth]{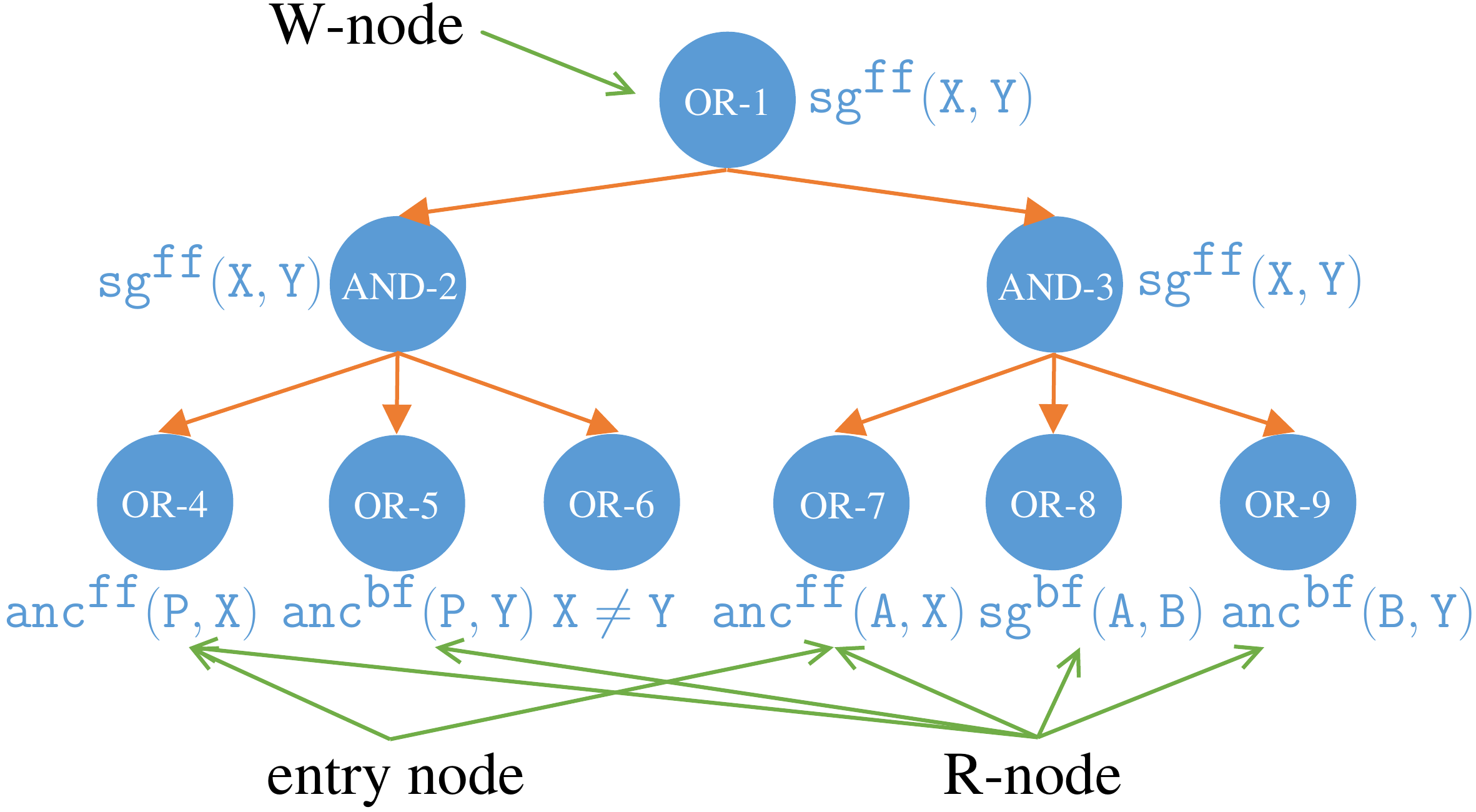}
\caption{AND/OR tree of \sg program in \cexp\ref{ex:sg}.}
\label{fig:sg-and-or}
\end{figure}

\begin{example}[AND/OR tree of \sg]\label{example:sg}
\cfig\ref{fig:sg-and-or} shows the adorned AND/OR tree of the same generation program \sg in \cexp\ref{ex:sg}, where (i) the text inside a node indicates its type and ID, e.g., ``\ornode{1}'' indicates that the root is an OR node with ID 1, and (ii) the text adjacent to a node shows the corresponding predicate with its adornment (${\tt b}$ or ${\tt f}$ in the $i$-th position means the $i$-th argument in a predicate $p$ is bound or free when $p$ is evaluated). Thus, \ornode{4}, \ornode{5}, \ornode{7}, \ornode{8}, and \ornode{9} are R-nodes, and \ornode{1} is a W-node. \ornode{4} and \ornode{7} are entry nodes in this program. Although \ornode{5} is an R-node, it is not an entry node since it is not the first R-node among its siblings. Similarly for \ornode{8} and \ornode{9}.

\end{example}

In the parallel evaluation of an AND/OR tree with one coordinator and $n$ workers, we create $n$ copies of the same AND/OR tree, and assign the $i$-th copy to the $i$-th worker. The evaluation is divided into $n$ disjoint parts, where the $i$-th worker evaluates an entry node by instantiating variables with constants from the $i$-th partition of the corresponding relation, while it has full access to all partitions of the corresponding relations for the remaining R-nodes. The parallel evaluation ensures the same workflow as the sequential pipelined evaluation by adding synchronization barriers in the nodes that represent recursion. For example, we create a synchronization barrier $B$, and add it to \ornode{1} of \cfig\ref{fig:sg-and-or} for every copy of the AND/OR tree. Now, the evaluation works as follows.
\begin{enumerate}
\item Each worker evaluates the exit rule by calling \gettuple{\andnode{2}} until it fails.  A worker waits at $B$ after it finishes.
\item Once all $n$ workers wait at $B$, the coordinator notifies each worker to start Step (3).
\item Each worker evaluates the recursive rule by calling \gettuple{\andnode{3}} until it fails. A worker waits at $B$ after it finishes.
\item Once all $n$ workers wait at $B$, the coordinator checks if there are new tuples derived in ${\tt sg}$. If so, the evaluation continues from Step (3); otherwise, the evaluation terminates.
\end{enumerate}

\subsection{Selecting a Parallel Plan}\label{sec:tree}
\bigdatalogmc uses a technique called {\em Read/Write Analysis} (RWA) \cite{DBLP:journals/vldb/YangSZ17} to help find the best discriminating sets to evaluate a program.
For a given set of discriminating sets, the 
RWA on an adorned AND/OR tree determines the actual program evaluation plan, including the type of lock needed for each derived relation, whether an OR node needs to acquire a lock before accessing the corresponding relation, and which partition of the relation an OR node needs to access when it accesses the relation through index lookups.
The analysis performs a depth-first traversal on the AND/OR tree that simulates the actual evaluation to check each read or write operation performed by the $i$-th worker. For each node $N$ encountered during the traversal, the following three cases are possible:
\begin{enumerate}
\item $N$ is an entry node. In this case, set it as the current entry node; then, for each W-node that is an ancestor of $N$ and is in the same stratum as $N$, determine whether the $i$-th worker only writes to the $i$-th partition of $R(p_w)$. This is done by checking if $p_e[\overline{X_j}] = p_w[\overline{X_k}]$,\footnote{For a predicate $p$, $R(p)$ denotes the relation that stores all tuples corresponding to facts about $p$; $p[\overline{X}]$ denotes a tuple of arity $|\overline{X}|$ by retrieving the arguments in $p$ whose positions belong to $\overline{X}$, and it is treated as a multiset of arguments when involved in equality checking.} where $p_e$ and $p_w$ are the predicates associated with $N$ and the W-node, respectively, and $\overline{X_j}$ and ${\overline{X_k}}$ are the corresponding discriminating sets.
\item $N$ is an R-node that reads from a derived relation. In this case, determine whether the $i$-th worker only reads from the $i$-th partition of $R(p_r)$ by checking if $\overline{X_k} \subseteq \overline{B}$ and $p_e[\overline{X_j}] = p_r[\overline{X_k}]$, where $p_e$ and $p_r$ are the predicates associated with the current entry node and $N$, respectively, $\overline{X_j}$ and ${\overline{X_k}}$ are the corresponding discriminating sets, and $\overline{B}$ is the set of positions for bound arguments in $N$.
\item $N$ is an R-node that reads from a base relation through a secondary index. In this case, determine whether the $i$-th worker only needs to read from one partition of $R(p_r)$ instead of all the partitions by checking if $\overline{X_k} \subseteq \overline{B}$, where $p_r$ is the predicate associated with $N$, ${\overline{X_k}}$ is the corresponding discriminating set, and $\overline{B}$ is the set of positions for bound arguments in $N$.
\end{enumerate}
We can formulate the problem of determining 
the best discriminating sets for a given program 
as an optimization problem that minimizes 
the {\em cost of program evaluation}. 
This is equivalent to minimizing the overhead of program evaluation over the ``ideal'' plan in which all the constraints are satisfied.
Now, for each OR node $N$ in the AND/OR tree, its contribution to the overhead of program evaluation is denoted by $c(N)$, and its value is heuristically set as follows:
\begin{displaymath}
c(N) = \left\{\begin{array}{ll}
3, & \mbox{if}~N~\mbox{needs to acquire an r-lock (read lock) before performing an} \\
& \mbox{index lookup and condition}~\overline{X_k} \subseteq \overline{B}~\mbox{is violated}; \\
1, & \mbox{if}~N~\mbox{needs to acquire a write lock before accessing the relation}; \\
0, & \mbox{otherwise}.
\end{array} \right.
\end{displaymath}
Thus, the optimization problem reduces to finding an assignment that minimizes $\sum_{N} c(N)$, where $N$ iterates over the set of OR nodes in the AND/OR tree. In \bigdatalogmc, this is achieved by enumerating all possible assignments using brute force, since the number of such valid assignments is totally tractable for most recursive queries of our interest. 
It is also important to take a closer look at the case where $c(N)$ equals three.
There are two parts in the corresponding condition: first, $N$ needs to acquire a read lock before performing an
 index lookup, and second, condition $\overline{X_k} \subseteq \overline{B}$ is violated. When $\overline{X_k} \subseteq \overline{B}$ is not true, this means we need to perform a lookup for each partition. This cost should be at least two, as there should be more than one partition during the parallel evaluation (otherwise, there is no need for parallelizing as there is only one partition and one processor). We also need to acquire a read lock for each lookup. However, we do not want to penalize this as much as acquiring a write lock, as acquiring a read lock is relatively less expensive. So the contribution from the read lock is counted as one, and the overall cost is summed as three.

\subsection{Experiments}

Now we introduce a set of experiments showcasing the performance of \bigdatalogmc compared to other (single and multicore) Datalog implementations, namely LogicBlox \cite{aref2015design}, DLV \cite{leone2006dlv}, {\sf clingo} \cite{gebser2014clingo}, and SociaLite \cite{seo2013socialite}. Additional experiments and details can be found in \cite{DBLP:journals/vldb/YangSZ17}.

\stitle{Configuration.}
We tested the performance of the above systems on a machine with four AMD Opteron 6376 CPUs (16 cores per CPU) and 256GB memory (configured into eight NUMA regions). The operating system was Ubuntu Linux 12.04 LTS. We used LogicBlox 4.1.9 and {\sc clingo} version 4.5.0. The version of DLV we used is a single-processor version~\footnote{The single-processor version of {\sc DLV} is downloaded from~
\url{http://www.dlvsystem.com/files/dlv.x86-64-linux-elf-static.bin}. Although a parallel version 
is available from \url{http://www.mat.unical.it/ricca/downloads/parallelground10.zip}, it is either much slower than the single-processor version, or it fails since it is a 32-bit executable that does not support more than 4GB memory required by evaluation.}, while for SociaLite we used the parallel version that was downloaded from~
\url{https://github.com/socialite-lang/socialite}.

\begin{figure}[t]
\centering
\includegraphics[
width=\textwidth]{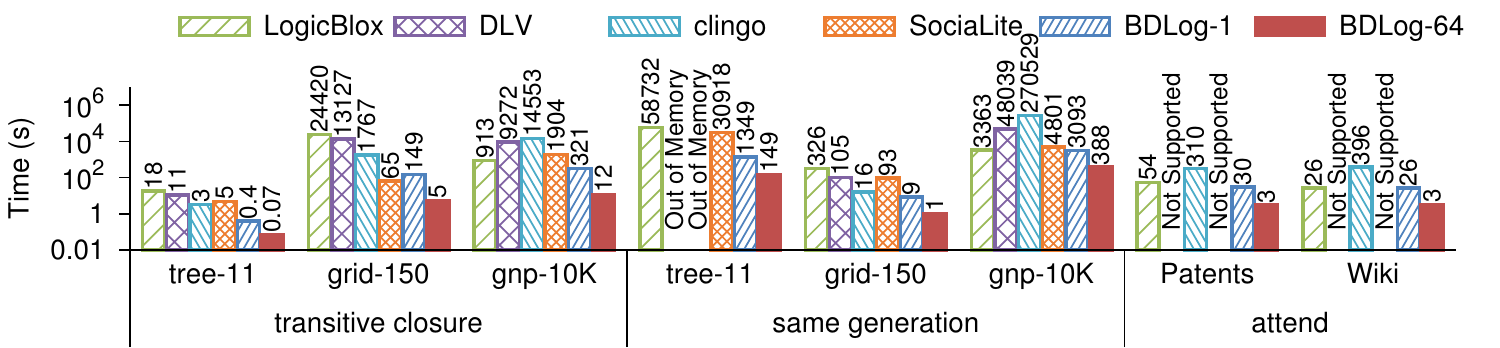}
\caption{Query evaluation time of recursive queries.}
\label{fig:recursive}
\end{figure}

\stitle{System Comparison.} \cfig\ref{fig:recursive} compares the evaluation time of the five systems on \tc, \sg, and \attend query. 
Bars for DLV and BDLog-1 show the evaluation time of {\sc DLV} and \bigdatalogmc using one processor, while bars for LogicBlox, Clingo, SociaLite, and BDLog-64 show the evaluation time of those systems over 64 processors. 
%
In our experiments, we observed both SociaLite and \bigdatalogmc had higher CPU utilization most of the time, as compared to LogicBlox and {\sc clingo}, with the latter utilizing only one processor most of the time.\footnote{These observations are obtained from the results of htop (see \url{https://hisham.hm/htop/}).}

When \bigdatalogmc is allowed to use only one processor, it always outperforms {\sc DLV} and {\sc clingo}. This comparison suggests that \bigdatalogmc provides a tighter implementation compared with the other two systems; specifically, we found that {\sc clingo}, although a multicore Datalog implementation, spends most of the time on the grounder that utilizes only one processor.

Moreover, with only one processor, \bigdatalogmc outperforms or is on par with LogicBlox and SociaLite, while LogicBlox and SociaLite are allowed to use all 64 processors. Naturally, \bigdatalogmc always significantly outperforms LogicBlox and SociaLite when it uses all 64 processors.
The performance gap between LogicBlox and \bigdatalogmc is largely due to the staged evaluation used by LogicBlox, which stores all the derived tuples in an intermediate relation, and performs deduplication or aggregation on the intermediate relation. For the evaluation that produces large amount duplicate tuples, such as ${\tt TC}$ on {\sf Grid150} and ${\tt SG}$ on {\sf Tree11}, this strategy incurs a high space overhead, and the time spent on the deduplication, which uses only one processor, dominates the evaluation time. 
SociaLite instead uses an array of hash tables with an initial capacity of around 1,000 entries for a derived relation, whereas \bigdatalogmc uses an append-only structure to store the tuples and a B+ tree to index the tuples. Although the cost of accessing a hash table is lower than that of a B+ tree, the design adopted by \bigdatalogmc allows a better memory allocation pattern as the relation grows. Such overhead is amplified when (i) multiple processors try to allocate memory at the same time, or (ii) the system has a high memory footprint.

Lastly, note that \bigdatalogmc
achieves a greater speedup (the speedup of BDLog-64 over BDLog-1) for ${\tt TC}$ than ${\tt SG}$ and ${\tt ATTEND}$ since no lock is used in ${\tt TC}$, while ${\tt SG}$ and ${\tt ATTEND}$ suffer from lock contention.

\section{Related Work}
\label{sec:related}

\stitle{Datalog Semantics.}
Supporting aggregates in recursion is an old and difficult problem which
has been the topic of much previous research work. Remarkably, previous approaches
had primarily focused on providing a formal semantics that could accommodate
the non-monotonic nature of the aggregates. In particular~\cite{DBLP:conf/vldb/MumickPR90}
discussed programs that are stratified w.r.t. aggregates operators
and proved that a perfect model exists for these programs. Then,
 \cite{DBLP:conf/slp/KempS91} defined extensions of the well-founded semantics
 to programs with aggregation, and later showed that programs with
 aggregates might have multiple and counter-intuitive stable models.
The notion of cost-monotonic  extrema aggregates was introduced by \cite{DBLP:journals/jcss/GangulyGZ95},
 using perfect models and well-founded
semantics, whereas ~\cite{DBLP:conf/pods/GrecoZG92} showed that
 their use to express greedy algorithms requires the  don't-care
 non-determinism of the stable-model semantics
provided by the choice construct.
An approach to  optimize  programs with extrema  was proposed by
\cite{Ganguly:1991:MMP:113413.113427}, and a general optimization technique
based on an early pruning of non-relevant facts
was proposed by  Sudarshan et al. \cite{DBLP:conf/vldb/SudarshanR91}.

A general approach to deal with the four aggregates, min, max, count and sum,
in a unified framework was  proposed by
\cite{ross1992monotonic} who advocated the use of
 semantics based on specialized lattices, different from
 set-containment, whereby each aggregate will then define a  monotonic
 mapping in its specialized lattice.
However several limitations of this proposal were
pointed out by Van Gelder~\cite{van1993foundations},
including the assumption that cost arguments of atoms are functionally
dependent from the other arguments. This is a property that does not
hold in many applications and it also difficult to determine,
since determining if a derived predicate satisfies a functional 
dependency is undecidable in general~\cite{DBLP:conf/sigmod/AbiteboulH88}.
In the following years, interest in aggregates for logic-based
systems focused on their use in the framework of  
Answer-Sets~\cite{DBLP:journals/aim/ErdemGL16}
which is less conducive to Big Data applications.
 
A renewed interest in Big Data analytics brought a revival of  Datalog
for  expressing more powerful data-intensive algorithms---including many that require aggregates in recursion. At
UCLA, researchers 
first introduced the notion of monotonic sum and count~\cite{mazuran2012extending,mazuran2013declarative}
and then proposed the comprehensive solution that is 
described in this paper and covers all four basic aggregates
along with efficient techniques for their efficient and scalable
implementation.

\stitle{Datalog Implementations.}
The Myria \cite{wang2015asynchronous} runtime supports Datalog evaluation using a pipelined, parallel, distributed execution engine that evaluates graph of operators.  Datasets are sharded and stored in PostgreSQL instances at worker nodes.  SociaLite~\cite{seo2013distributed} is a Datalog language implementation for social network analysis. SociaLite programs are evaluated by parallel workers that use message passing to communicate. Both SociaLite and Myria support aggregation inside recursion focusing on their operational semantics. The 
 lattice-based  approach of Ross and Sagiv \cite{ross1992monotonic} is proposed as the possible
 basis for  a declarative semantics,  but  no approach on how to overcome its  limitations is discussed.
Furthermore, the advent of graphics processing units (GPUs) has recently led to Datalog implementations on GPUs for relational learning algorithms \cite{reviewer3-relationalGPU}. Since the transfer of data  between host and GPU memory incurs in significant cost, Datalog implementations on GPUs \cite{reviewer1-datalogGPU} optimize this cost through efficient memory management schemes. 

\vspace{1ex}
\stitle{Parallel Datalog Evaluation and Languages.}  Previous research on parallel evaluation of Datalog programs determined that some programs are  evaluable in parallel without redundancy and without processor communication or synchronization \cite{DBLP:conf/sigmod/WolfsonS88}. Such programs are called decomposable.
Our parallel implementations identify decomposable programs via syntactic analysis of program rules using the generalized pivoting method \cite{generalizedpivoting}. 
Others have also explored the idea of applying extensions of simple 0-1 laws on Datalog programs to derive at a parallelization plan that maximizes the expected performance \cite{reviewer2-probView}. 

Many works produced over twenty years ago focused on parallelization of bottom-up evaluation of Datalog programs \cite{zhang1995data}, however they were largely of a theoretical nature. For instance \cite{van1993foundations} proposed a message passing framework for parallel evaluation of logic programs.  Techniques to partition program evaluation efficiently among processors \cite{DBLP:conf/sigmod/WolfsonO90}, the trade-off between redundant evaluation and communication \cite{DBLP:conf/sigmod/GangulyST90,DBLP:journals/jlp/GangulyST92}, and classifying how certain types of Datalog\ programs can be evaluated \cite{Cohen:1989:WSP:73721.73742} were also studied.  
A parallel semi-na{\"i}ve fixpoint has been proposed for message passing \cite{DBLP:conf/sigmod/WolfsonO90} that includes a step for sending and receiving tuples from other processors during computation.  The PSN used in this work applies the same program over different partitions and shuffle operators in place of processor communication.
Parallel processing of recursive queries in particular is also a  well-studied problem. One such example is \cite{bell1991pipelined}, where the recursive query is first transformed into a canonical form and then evaluated in a pipelined fashion. 
%

Recently, Semantic Web reasoning systems dealing with RDF data has utilized this early research in parallel implementations of semi-na{\"i}ve evaluation \cite{Abiteboul:1995:FDL:551350} to handle recursive Datalog rules much like commercial systems as LogicBlox. One such system is RDFox \cite{reviewer1-parallelMatl} which is a main-memory, multi-core RDF system that uses a specialized RDF indexing data structure to ensure largely lock-free concurrent updates. 
It is also important to mention in this regard that, with the emergence of large Knowledge Graphs \cite{reviewer1-knowledgeGraph}, the Semantic Web community has significantly contributed to the ongoing research in Datalog reasoning. In fact, many reasoning systems encode RDF data, as represented in Knowledge Graphs, into ternary database predicates for writing elegant Datalog rules, which in turn, have to be efficiently evaluated. One such recent system is Vlog \cite{reviewer1-knowledgeGraph} which exploits column-based memory layout along with selective caching of certain subquery results. However, Vlog is intrinsically sequential in nature and does not have a parallel or distributed implementation.

Among the distributed Datalog\ languages, it is noteworthy to mention \textit{OverLog} \cite{declarativeoverlays,evitaraced}, used in the P2 system to express overlay networks, and \textit{NDlog} \cite{declarativenetworking} for declarative networking. The \textit{Bloom$^{L}$} \cite{blooml} distributed programming language uses various monotonic lattices, also based on the semantics of \cite{ross1992monotonic}, to identify program elements not requiring coordination.  \cite{scalingdatalog} showed how XY-stratified Datalog can support computational models for large-scale machine learning, although no full Datalog language implementation on a large-scale system was provided.  

 \stitle{Beyond Datalog: Parallel Execution of Logic Programs.}
In logic programming, programs are evaluated in a top-down fashion through \emph{unification}.
An extensive body of research was produced  on parallel logic programming, dating  back to 1981~\cite{Gupta:2001:PEP:504083.504085,deKergommeaux:1994:PLP:185403.185453}).
Two major approaches exists for parallelizing logic programs: the \emph{implicit} approach assumes that the framework is able to parallelize the given input program automatically without any programmer intervention. Conversely, in the \emph{explicit} case specific constructs are introduced into the source language to guide the parallel evaluation.
The approach used in our \bigdatalog systems is implicit parallelism where any input Datalog program is automatically parallelized by the runtime.

In implicit parallel logic programming, three main forms of parallelism exists: (i) \emph{And-Parallelism} whereby multiple literals are evaluated concurrently; (2) \emph{Or-Parallelism} where instead clauses are evaluated in parallel; and (3) \emph{Unification Parallelism} in which the unification process is parallelized. 
Our parallel evaluation of Datalog programs is a form of Or-Parallelism where data is partitioned such that different rule instantiations are evaluated concurrently.

It is also important to note that the growth of Semantic Web data also propelled considerable research on large-scale reasoning on distributed frameworks like MapReduce~\cite{mapreduce}. One such example is the WebPIE system \cite{reviewer1-webpie}  that implements forward reasoning for RDFS over MapReduce framework.
The key ideas originating from distributed MapReduce frameworks used for Semantic Web reasoning were also applied for description logic $\mathcal{EL}^+$ \cite{reviewer1-mapReduce} for $\mathcal{EL}^+$\ ontology classifications. 
In this era of big data, the Semantic Web community also led considerable research efforts towards large-scale non-monotonic reasoning of RDF data. One such paper is \cite{reviewer1-defeasibleReasoning} which proposed a MapReduce based parallel framework for defeasible logic and predicates of any arity in presence of noisy data. 
In the same vein of large scale non-monotonic reasoning, the authors of \cite{reviewer1-TPLP} proposed a similar data parallel MapReduce framework for well-founded semantics computation through efficient implementations of joins and anti-joins. 

%
%
%
%
\section{Conclusion}\label{sec:conclusion}
By embracing the Horn-clause logic of Prolog but not its 
operational constructs such as the cut, 
Datalog researchers, 30 years ago, embarked in a significant expedition 
toward declarative languages in which logic alone rather than
Logic+Control \cite{DBLP:journals/cacm/Kowalski79} can be
used to specify algorithms. 
Significant progress toward this ambitious goal was made in the 
90s with  techniques such as semi-na{\"i}ve  fixpoint and 
magic sets that support recursive
Datalog programs by bottom-up computation and 
implementation techniques from relational DB systems. 
As described in  Section \ref{sec:related}, however,
 declarative semantics for algorithms that
 require aggregates in recursion largely
remained an unsolved problem for this first generation of deductive DB systems. 
Moreover,  
Datalog scalability via parallelization was only discussed in papers,
until recently when the availability of
new parallel platforms and an explosion of
interest in BigData  renewed  interest in Datalog and 
its parallel implementations on multicore and distributed systems.

In this paper, we have 
provided an in-depth description of the UCLA'a BigDatalog/DeAL project that
is of significance because of its  (i)  historical continuity with  first-generation  Datalog systems ($\cal LDL{++}$ was supported and extended at UCLA for several years \cite{ldl++}), (ii) 
 implementation on multiple platforms, with levels of performance
that surpass those of competing Datalog systems, GraphX applications, and
even Apache Spark applications written in Scala, and (iii)  support for a wide range of declarative algorithms using the rigorous non-monotonic semantics for recursive programs with aggregates introduced in \cite{DBLP:journals/tplp/ZanioloYDSCI17}.


Furthermore, we believe that the  use
of aggregates in recursive rules made possible by 
\prem~\cite{DBLP:journals/tplp/ZanioloYDSCI17} 
can lead to beneficial extensions in several application 
areas, e.g., KDD algorithms, and  in related 
logic-based systems, including, e.g., those that use 
tabled logic programming~\cite{DBLP:journals/tplp/SwiftW12} 
and  Answer Sets~\cite{DBLP:journals/aim/ErdemGL16}. Therefore we
see many interesting  new topics deserving further investigation, suggesting that
 logic and databases remains a vibrant research area~\cite{DBLP:conf/amw/ZanioloYIDSC18} although many years have  passed since it  
was first introduced~\cite{MinkerSZ14}.

\section*{Acknowledgements}

We  would  like  to  thank  the  reviewers  for  the  comments  and  suggested  improvements.  This
work was supported in part by NSF grants IIS-1218471, IIS-1302698 and CNS-1351047, and
U54EB020404 awarded by NIH Big Data to Knowledge (BD2K).
%
\bibliographystyle{acmtrans}
\bibliography{tplp}
\end{document}